\documentclass{ws-mplb}
\usepackage{amsfonts,amssymb,amsmath}
\usepackage{graphicx}
\usepackage{mathrsfs}
\usepackage{amsmath}
\usepackage[super,sort,compress]{cite} 

\newcommand{\onlinecite}[1]{\hspace{-1 ex} \nocite{#1}\citenum{#1}} 

\makeatletter
	\renewcommand{\theequation}{
	\thesection.\arabic{equation}}
	\@addtoreset{equation}{section}

	\renewcommand{\thefigure}{
	\thesection.\arabic{figure}}
	\@addtoreset{figure}{section}

	\renewcommand{\thetable}{
	\thesection.\arabic{table}}
	\@addtoreset{table}{section}
\makeatother

\begin{document}

\markboth{H. Koizumi}{Instructions for typing manuscripts (paper's title)}

%
\catchline{}{}{}{}{}
%

\title{Theory of Supercurrent in Superconductors}

\author{\footnotesize Hiroyasu Koizumi}

\address{Division of Quantum Condensed Matter Physics, Center for Computational Sciences, University of Tsukuba\\
Tsukuba, Ibaraki 305-8577, Japan
\\
koizumi.hiroyasu.fn@u.tsukuba.ac.jp}

\author{Alto Ishikawa}
\address{Graduate School of Pure and Applied Sciences,  University of Tsukuba \\ Tsukuba, Ibaraki 305-8573, Japan
\\
s2020321@s.tsukuba.ac.jp}

\maketitle

\begin{history}
\received{(Day Month Year)}
\revised{(Day Month Year)}
\end{history}

\begin{abstract}
In the standard theory of superconductivity, the origin of superconductivity is the electron-pairing.
The induced current by a magnetic field is calculated by the linear response to the vector potential, and the supercurrent is identified as the dissipationless flow of the paired-electrons, while single electrons flow with dissipation.
This supercurrent description suffers from the following serious problems: 1) it contradicts the reversible superconducting-normal phase transition in a magnetic field observed in type I superconductors; 2) the gauge invariance of the supercurrent induced by a magnetic field requires the breakdown of the global $U(1)$ gauge invariance, or the non-conservation of the particle number; 3) the explanation of the ac Josephson effect is based on the boundary condition that is different from the real experimental one.

We will show that above problems are resolved if the supercurrent is attributed to the collective mode arising from the Berry connection for many-body wave functions. 
The problem 1) is resolved by attributing the appearance and disappearance of the supercurrent to the abrupt appearance and disappearance of topologically-protected loop currents produced by the Berry connection; the problem 2) is resolved by assigning the non-conserved number to that for the particle number participating in the collective mode produced by the Berry connection; and the problem 3) is resolved by identifying the relevant phase in the Josephson effect is that arising from the Berry connection, and using the modified Bogoliubov transformation that conserves the particle number.

We argue that the required Berry connection arises from spin-twisting itinerant motion of electrons. For this motion to happen, the Rashba spin-orbit interaction has to be added to the Hamiltonian for superconducting systems.. 
The collective mode from the Berry connections is stabilized by the pairing interaction that changes the number of particles participating in it; thus, the superconducting transition temperatures for some superconductors is given by the pairing energy gap formation temperature as explained in the BCS theory.
The topologically-protected loop currents in this case are generated as cyclotron motion of electrons that is quantized by the Berry connection even without an external magnetic field.

We also explain a way to obtain the Berry connection from spin-twisting itinerant motion of electrons for a two-dimensionL model where
the on-site Coulomb repulsion is large and doped holes form small polarons. In this model, the electron-pairing is not required for the stabilization of the collective mode, and the supercurrent is given as topologically-protected spin-vortex-induced loop currents (SVILCs). 

\end{abstract}

\keywords{superconductivity; supercurrent; Berry connection; Rashba spin-orbit interaction}

\section{Introduction}
Since the discovery of high temperature superconductivity in cuprates, its mechanism has been a focus of attention in condensed matter physics and materials science \cite{Muller1986}.
The cuprate superconductivity exhibits a number of differences compared with the superconductivity explained by the standard theory based on the BCS one \cite{BCS1957}. Of all differences the most peculiar is the existence of the pseudogap phase above the superconducting phase in the $p$-$T$ plane, where $p$ is the hole doping per Cu in the CuO$_2$ plane of the cuprate and $T$ the temperature. It appears that the d-wave pairing gap starts to exist in the pseudogap phase but the system is not superconducting; this suggests that the identification of the existence of the pairing energy gap to the  superconducting state may not be valid.
In spite of tremendous efforts, no widely-accepted theory for the cuprate superconductivity exists at present. It is now clear that a marked departure from the standard theory is necessary for the elucidation of the cuprate superconductivity.

Meanwhile  the completeness of the BCS theory as a theory for superconductivity has been questioned. The question is on the generation of the supercurrent: does the BCS theory really explain the supercurrent generation in superconductors?
There are three points that indicate the answer is negative.  

\begin{figure}
\begin{center}
\includegraphics[scale=0.5]{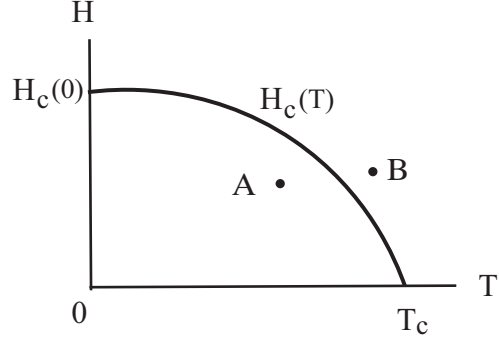}
\end{center}
\caption{The phase diagram for type I superconductor in $T$-$H$ plane, where $T$ is the temperature and $H$ is the applied magnetic field. In the figure, ``A'' is in the superconducting phase and
``B'' is in the normal phase. They are thermodynamical equilibrium states, thus, the change of states between them is reversible irrespective of the path taken in the $T$-$H$ plane in an ideal situation.}
\label{T-H}
\end{figure}

One of them is concerning the superconducting-normal phase transition in a magnetic field raised by Hirsch \cite{Hirsch2017}: the supercurrent generation mechanism of the standard theory cannot explain the reversible superconductor-normal transition in the $H$-$T$ plane of type I superconductors, where $H$ is the external magnetic field.

The Meissner effect \cite{Meissner1933} indicates that the superconducting-normal phase transition in a magnetic field in this plane is reversible. For example, the change of states from A to B in Fig.~\ref{T-H} is reversible. The reversibility was confirmed through a series of work \cite{Keesom1,Keesom2,Keesom3, Keesom4,Keesom}, and the state of the art calorimetry indicates that 99.99\% of the supercurrent stops without current carriers undergoing irreversible collisions (see Appendix B of Ref.~\onlinecite{Hirsch2017}). 

On the other hand, in the standard theory, the supercurrent generated by the dissipationless flow of electron pairs becomes the dissipative one 
during the superconducting to normal phase transition due to the appearance of broken pairs that flow with dissipation \cite{Hirsch2017,Hirsch2018,Hirsch2020}. The Joule heat generated by the flow of single electrons during the transition will make it irreversible one.

Another problem is concerning the breakdown of the global $U(1)$ gauge invariance. It means that the number of particles in the superconductor is fluctuating, or the number of electrons in the superconductor is not fixed.  
In other words, during the phase transition from B to A in Fig.~\ref{T-H}, somehow the number of electrons start to fluctuate during the crossing of the $H_c(T)$ curve. 

The BCS theory uses the variational wave function that is a linear combination of different particle number states \cite{BCS1957}. The original intention to use such a variational state was to facilitate calculations involving the electron pair formation \cite{BCS1957}, and it has been argued that it should be considered in such a way \cite{Peierls1991}. 
However, it is taken as a pure state in the standard theory. Further, it has been argued that this breakdown of the global $U(1)$ gauge invariance is the hallmark of the superfluid systems \cite{Anderson66}. 

The global $U(1)$ breaking formalism was first introduced by Bogoliubov in the weakly-interacting boson system \cite{Bogolubov47}, where
the canonical transformation that mixes different particle number states was used. By this canonical transformation (Bogoliubov transformation), he obtained the excitation spectrum that satisfies the Landau's criterion for the occurrence of superfluidity \cite{Landau}. He also showed that such a transformation is possible for the superconducting state of the BCS theory and enables to obtain the BCS results more easily than the method employed by BCS \cite{Bogoliubov58}. Thereby, the Bogoliubov's canonical transformation formalism has been proven to be a successful calculation tool for both bosonic superfluid systems and fermionic ones.
Now the extension of the Bogoliubov's treatment of the superconductivity, the Bogoliubov-de Gennes method is a common tool to
study superconducting systems \cite{deGennes}.

In spite of the success of the Bogoliubov transformation, the assumed breakdown of the global $U(1)$ invariance has been claimed to be impossible since the relevant Hamiltonian conserves the particle number \cite{WWW1970,Peierls1991,Peierls92,Leggett2001,LeggettBook,Leggett2016}. Such a pure state of a linear combination of different particle number states  may be prepared at a time in principle, but it will soon be collapsed to a mixed state of different particle number states by the decoherence caused by the interaction with environment \cite{Zurek2002}. 

Recently, Leggett showed that the excitation spectrum obtained by Bogoliubov can be obtained without using the breakdown of the $U(1)$ gauge invariance \cite{Leggett2001,LeggettBook}. Furthermore, the Bogoliubov transformation used for superconductivity problems can be cast in the particle number fixed form \cite{koizumi2019}.
Actually, it is argued that the fluctuation of the number of particles in the original Bogoliubov's treatment should be re-interpreted as that for the particles participating in the condensate part. 

In the standard theory, the breakdown of the global $U(1)$ invariance plays an essential role in the gauge invariance of the induced current calculated by the linear response theory \cite{Anderson1958a,Anderson1958b,Anderson66,Nambu1960}. Actually, the use of the liner response theory is also problematic from the view point of the Joule heating problem mentioned above;
the current calculated by the linear response theory will obey the fluctuation-dissipation relation, yielding Joule heating from the flow of broken electron-pairs, thus contradicts the reversible superconducting-normal phase transition in a magnetic field.

\begin{figure}
\begin{center}
\includegraphics[scale=0.6]{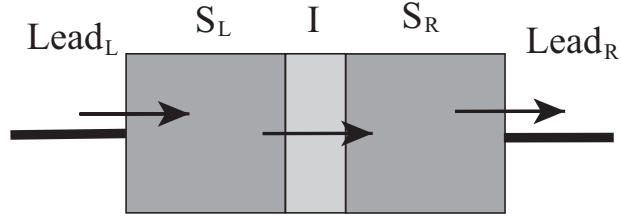}
\end{center}
\caption{Josephson junction S$_{\rm L}$-I-S$_{\rm R}$ (superconductor-insulator-superconductor junction) connected to leads, denoted as Lead$_{\rm L}$ and Lead$_{\rm R}$. Josephson included only S$_{\rm L}$$\rightarrow$S$_{\rm R}$ electron transfer.
In the real experimental situation, however, electron transfers from Lead$_{\rm L}$$\rightarrow$ S$_{\rm L}$ and S$_{\rm R}$$\rightarrow$ Lead$_{\rm R}$ also occur. It has been shown that the latter electron transfers also contribute to the time-variation of the order parameter phase \cite{Koizumi2011,HKoizumi2015}.}
\label{Josephson}
\end{figure}

\begin{figure}
\begin{center}
\includegraphics[scale=0.55]{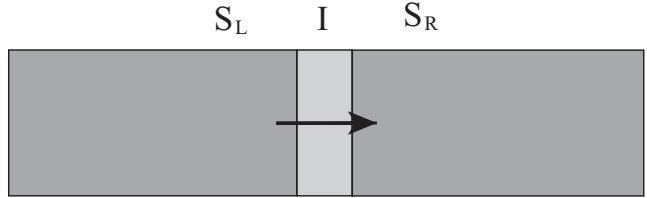}
\end{center}
\caption{Josephson tunneling considered in the Josephson's derivation.}
\label{Josephson2}
\end{figure}

Now we move to the last problem. It  is concerning the ac Josephson effect \cite{Josephson62}. The experimental set-up for the ac Josephson effect is depicted in Fig.~\ref{Josephson}. The boundary condition adopted by Josephson is given in Fig.~\ref{Josephson2}.
The boundary condition by Josephson and the real experimental one are actually different \cite{Koizumi2011,HKoizumi2015}. 
The Josephson's derivation considered only the tunneling between the two superconductors in the junction; however, in the real experimental situation, the dc current is supplied through the leads connected to the junction, and the voltage appears when a radiation field is present under the existence of the dc current.  
 
By adopting the experimental boundary condition\cite{Koizumi2011,HKoizumi2015}, it is shown that the observed Josephson relation
\begin{eqnarray}
\dot{\phi}={{2eV} \over \hbar}
\label{eqJR}
\end{eqnarray}
 where $\phi$ is related to the current through the Josephson junction as $J=J_c \sin \phi$ ($J_c$ is a parameter for the junction), means that the charge of the supercurrent carrier is $-e$; there are two contributions to $\dot{\phi}$, one from the electric field across the two superconductors
 (denoted by the arrow from S$_{\rm L}$ to S$_{\rm R}$ in Fig.~\ref{Josephson}, which is the one included in the Josephson's derivation), and another from the chemical potential difference brought-in when particles enter from one lead to the junction and exit to the other lead from the junction (denoted by the arrows from Lead$_{\rm L}$ to S$_{\rm L}$ and S$_{\rm L}$ to Lead$_{\rm R}$ in Fig.~\ref{Josephson}). 
 These two contributions are physically different but the same in magnitude due to the balance between the chemical potential difference and the voltage from the electric field. The factor $2$ in Eq.~(\ref{eqJR}) arises from the fact that there are two contributions with the same magnitude.
 It is often likened that superconductivity is a Bose-Einstein condensation of electron pairs by regarding electron pairs as bosons \cite{FeynmanText1}, and the Josephson relation is considered as a manifestation of $2e$ charge on the charge carrier; however, it is not so.
 
In his derivation, Josephson employed the original Bogoliubov transformation;
in this case, the supercurrent flow through the junction has to be the current flow with the charge $-2e$ carriers in order to avoid the electron-pair breaking. 
However, when the particle number conserved  Bogoliubov transformation is used, the supercurrent flow with the charge $-e$ carrier is possible \cite{koizumi2019,koizumi2020}. Then, the observed Josephson relation in Eq.~(\ref{eqJR}) is obtained by employing the boundary condition of the experiment without breaking the global $U(1)$ gauge invariance. 
 
The purpose of this review is to present a new supercurrent generation mechanism that resolves the above three problems. \cite{koizumi2019,koizumi2020}. It utilizes the Berry phase \cite{Berry} that was not known when the BCS paper appeared, but becomes now one of the basic ingredients of many-body physics. 
 We argue that supercurrent is created as a collection of topologically-protected loop currents produced by the Berry connection, where the ``topological protection" means the disappearance or  appearance of the supercurrent is accompanied by the change of the topological quantum numbers associated with the Berry connection, which gives rise to a certain rigidity to the stability of the current. 
The required nontrivial Berry connection is argue to arise from spin-twisting itinerant motion of electrons, where $\pi$-flux Dirac strings exist at the centers of the spin-twisting \cite{koizumi2020}.

It is very interesting to note that the present theory is somewhat similar to the molecular vortex theory of Maxwell \cite{Maxwell1,Maxwell2,Maxwell3,Maxwell4}.
The nontrivial Berry connection in the present theory may be likened to the ``idle wheel" envisaged by Maxwell in his loop current generation mechanism \cite{Maxwell2}. 

\section{London's theory of superconductivity with superpotential $\chi_s$}
\label{Section2}

We shall start with reexamining the London's theory of superconductivity \cite{London1950}.

He proposed that the supercurrent in a magnetic field is given by
\begin{eqnarray}
{\bf J}=-{ {n e^2} \over {m}}{\bf A}
\label{eqLondon0}
\end{eqnarray}
where ${\bf A}$ is the vector potential, ${\bf J}$ the current density, $n$ the superconducting electron density, and $m$ the mass.
Note that since ${\bf J}$  is a  physical entity, ${\bf A}$ must be also a physical entity, which is in accordance with the Aharonov and Bohm effect that indicates the gauge potential contains physics that cannot be described by 
the electric field ${\bf E}$ and the magnetic field ${\bf B}$ \cite{AB1959} confirmed by Tonomura el al.\cite{Tonomura1986}

In addition to the above formula, the London's theory of superconductivity actually contains the ``superpotential'' $\chi_s$ ($\chi$ in  Ref.~\onlinecite{London1950}) that gives rise to a long range order of the average momentum given by
\begin{eqnarray}
{\bf p}_s=\nabla \chi_s
\end{eqnarray}

Including $\chi_s$, Eq.~(\ref{eqLondon0}) is written as
\begin{eqnarray}
{\bf J}=-{ {ne^2} \over {m}}\left({\bf A}^{\rm em} +e^{-1}\nabla \chi_s \right)
\label{eqLondon01}
\end{eqnarray}
where ${\bf A}^{\rm em}$ is the ordinary electromagnetic vector potential.
It is assumed that 
\begin{eqnarray}
 {\bf A}^{\rm em} +e^{-1}\nabla \chi_s
\end{eqnarray}
is gauge invariant \cite{London1950}.

 \begin{figure}
\begin{center}
\includegraphics[scale=0.55]{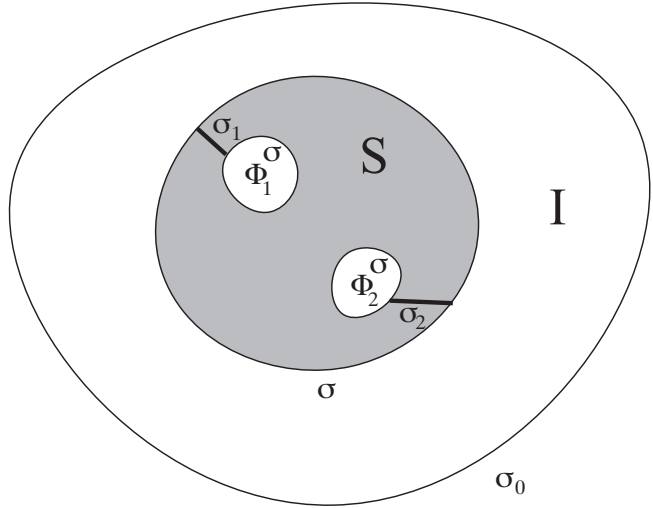}
\end{center}
\caption{Superconductor ``S'' with $k$ holes ($k=2$) surrounded by insulator ``I''. $\sigma$ is the surface of the superconductor, and $\sigma_0$ is the surface of the whole system. Fluxoid for the $k$th hole is denoted by $\Phi_k$. The $(k+1)$-fold connected region of the superconductor becomes a singlely-connected one by inserting $k$ dividing surfaces denoted by $\sigma_k$.}
\label{London-holes}
\end{figure}

The angular variable $\chi_s$ is allowed to be multi-valued function of the coordinate.
If the superconductor is a $(k+1)$-fold connected one (see Fig.~\ref{London-holes} for a $k=2$ case), the multi-valuedness of $\chi_s$ allows the presence of $k$ fluxoids given by
\begin{eqnarray}
 \Phi_k=\oint_{(k)} e^{-1}\nabla \chi_s \cdot d{\bf r}
 \label{eqFluxoid}
\end{eqnarray}
where the integration is performed along a loop encircling the $k$th hole. In quantum theory, the value of $\Phi_k$ is quantized.

London showed that the current distribution and magnetic field in the superconductor are uniquely given by $\Phi_k$ and the magnetic field at the surface $\sigma_0$ in Fig.~\ref{London-holes} \cite{London1950}. In other words, the mechanism of the supercurrent generation is given if the origin of $\nabla \chi_s$ and values of fluxoids $\Phi_k$ are explained.
It is also important to note that since $\Phi_k$ has a quantized value, its change is abrupt and discrete; thus, the changes of the current distribution and magnetic field in the superconductor caused by the change of $\Phi_k$ are abrupt and discrete.

Later, we attribute the origin of $\nabla \chi_s$ to the non-trivial Berry connection arising from the spin-twisting itinerant motion of electrons. The single-valued requirement of the wave function with respect to the electron coordinates gives rise to the condition for the quantized values for $\Phi_k$, where ``holes'' are centers of the spin-twisting or the zeros of the wave function. 

\section{Gauge potential, gauge transformation, and London formula}

Let us examine the London formula in Eq.~(\ref{eqLondon0}) in this section.

The Schr\"{o}dinger equation for the charged particle with charge $q$ is invariant by the following transformation for the wave function $\psi$
\begin{eqnarray}
\psi'=\exp \left( i { q \over {\hbar }} f \right) \psi
\label{eqWave}
\end{eqnarray}
where $f$ is a function, 
and the $U(1)$ gauge potential, 
\begin{eqnarray}
{\bf A}'={\bf A}+\nabla f, \quad  \varphi'=\varphi-{ {\partial f} \over {\partial t}}
\label{eqVector}
\end{eqnarray}
where ${\bf A}$ and $\varphi$ are the vector and scalar potentials, respectively.
The degree of freedom in the choice of the gauge potential by the above transformation is the gauge degree of freedom.

If the number of particles is $N$, Eq.~(\ref{eqWave}) becomes,
\begin{eqnarray}
\Psi'({\bf r}_1, \cdots, {\bf r}_N)=\exp \left( i { q \over {\hbar }}  \sum_{j=1}^N f ({\bf r}_j)\right) \Psi ({\bf r}_1, \cdots, {\bf r}_N)
\label{eqWave2}
\end{eqnarray}
where ${\bf r}_j$ is the coordinate of the $j$th particle.

It should be noted that the phase factor in Eq.~(\ref{eqWave2})
\begin{eqnarray}
\exp \left( i { q \over {\hbar }}  \sum_{j=1}^N f ({\bf r}_j)\right) 
\end{eqnarray}
gives rise to the whole system motion; thus, the gauge transformation is tied to the modification of the whole system motion \cite{FluxRule}.

We may choose the Coulomb gauge
\begin{eqnarray}
\nabla \cdot {\bf A}=0
\label{eqCoulomb}
\end{eqnarray}
This is the gauge adopted in the BCS theory \cite{BCS1957}.

The equations for ${\bf A}$ and $\varphi$ in this gauge are given by
\begin{eqnarray}
\left( \nabla^2 - { 1 \over c^2} { {\partial^2} \over {\partial t^2}}\right) {\bf A}- { 1 \over c^2} \nabla{ {\partial \varphi} \over {\partial t}}=-\mu_0 {\bf J} 
, \quad \nabla^2\varphi=-{ 1 \over {\epsilon_0}}\rho
\end{eqnarray}
where $\mu_0$ is the vacuum permeability and $\epsilon_0$ is the vacuum permittivity; the speed of light $c$ is given by $c={1 \over \sqrt{\mu_0\epsilon_0}}$.

From the above equations and using the condition in Eq.~(\ref{eqCoulomb}), the conservation of the charge 
\begin{eqnarray}
\nabla \cdot  {\bf J} + { {\partial} \over {\partial t}}\rho=0
\label{eq22}
\end{eqnarray}
is obtained, where $\rho$ is the charge density.

The gauge condition in Eq.~(\ref{eqCoulomb}) still leaves the following change,
\begin{eqnarray}
{\bf A} \rightarrow {\bf A}+ \nabla f, \quad \nabla^2 f=0
\label{eqFixchi}
\end{eqnarray}
where $f$ is a function.

For example, we can perform
\begin{eqnarray}
{\bf A} \rightarrow {\bf A}+ {\bf A}_0, \quad {\bf A}_0 \mbox{ is a constant}
\end{eqnarray}
with the accompanying change in the wave function
\begin{eqnarray}
\psi \rightarrow \exp \left( i { q \over {\hbar }} {\bf A}_0 \cdot {\bf r} \right) \psi
\end{eqnarray}

The right-hand-side of Eq.~(\ref{eqLondon0}) changes by the gauge transformation in Eq.~(\ref{eqFixchi}).
On the other hand, the left-hand-side remains the same. This is an apparent discrepancy.  

The resolution of this discrepancy is achieved in the standard theory as follows.
The standard theory is based on the BCS theory\cite{BCS1957}. It employs the particle number non-conserving state vector
\begin{eqnarray}
|{\rm BCS} \rangle=\prod_{\bf k}\left(u_{\bf k}+v_{\bf k}c^{\dagger}_{{\bf k} \uparrow}c^{\dagger}_{-{\bf k} \downarrow}
 \right)|{\rm vac} \rangle
\label{BCS0}
\end{eqnarray}
where $c^{\dagger}_{{\bf k} \sigma}$ is the electron creation operator with wave vector ${\bf k}$ and spn $\sigma$, and real parameters $u_{\bf k}$ and $v_{\bf k}$ satisfy $u_{\bf k}^2+v_{\bf k}^2=1$. This state vector is very convenient in calculating 
the electron pairing state.

In the original BCS paper, the induced current by the magnetic field was calculated using the perturbation theory (equivalent to the linear response theory). It is given by
\begin{eqnarray}
\lim_{{\bf q} \rightarrow 0} {\bf j}({\bf q})=-{{\Lambda} \over {\Lambda}_T}{ {ne^2} \over {m}}{\bf a}({\bf q})
\label{eqLondon2}
\end{eqnarray}
with 
\begin{eqnarray}
\lim_{T \rightarrow 0} {{\Lambda} \over {\Lambda}_T}=1,  \quad \lim_{T \rightarrow T_c} {{\Lambda} \over {\Lambda}_T}=0
\end{eqnarray}
where ${\bf j}({\bf q})$ and ${\bf a}({\bf q})$ are Fourier components of ${\bf J}$ and ${\bf A}$. This corresponds to Eq.~(\ref{eqLondon0}).

The gauge in Eq.~(\ref{eqCoulomb}) was adopted with the supplementary condition
\begin{eqnarray}
  {\bf A}=0 \ \mbox{    if the magnetic field is zero}
\end{eqnarray}

The right-hand-side of Eq.~(\ref{eqLondon2}) still depends on the choice of the gauge. We may add $\nabla f$ to ${\bf A}$ 
by the gauge transformation in Eq.~(\ref{eqFixchi}), and if it is done, ${\bf j}({\bf q})$ becomes different. This is the discrepancy.

In the standard theory, a resolution of this problem was achieved using the fact that $|{\rm BCS} \rangle$ breaks the global $U(1)$ invariance. First, the relation in Eq.~(\ref{eqLondon2}) is expressed using a kernel $K_{\mu \nu}(q)$,
\begin{eqnarray}
j_{\mu}(q)=-\sum_{\nu}  K_{\mu \nu}(q) a_{\nu} (q)
\end{eqnarray}
where $j_\mu$ is the four-component current density vector, and $a_{\nu} (q)$ is the four-component gauge potential vector; $q$ is the four-component wave vector.

The gauge invariance with respect to the following gauge transformation
\begin{eqnarray}
a_{\nu} (q) \rightarrow a_{\nu} (q) +iq_{\nu}f(q)
\end{eqnarray}
is given by
\begin{eqnarray}
\sum_{\nu} K_{\mu \nu}(q) q_{\nu} =0
\label{eqKernel}
\end{eqnarray}
This can be shown to be satisfied using the breakdown of the global $U(1)$ gauge invariance in the BCS superconducting state \cite{Anderson1958b,Nambu1960,SuperBook}. 
Particularly, Nambu used the Bogoliubov transformation and the Ward-Takahashi identity arising from the conservation of the charge \cite{Ward,Takahashi57}, found the existence of the boson mode, the {\em Nambu-Goldstone mode}, that guarantees the gauge invariance of the induced current. Thus, the breakdown of the global $U(1)$ gauge invariance is an essential ingredient of the standard theory.

For the global $U(1)$ gauge change 
\begin{eqnarray}
c^{\dagger}_{{\bf k} \sigma} \rightarrow e^{{ i \over 2} \theta}c^{\dagger}_{{\bf k} \sigma}, \quad c_{{\bf k} \sigma} \rightarrow e^{-{ i \over 2} \theta}c_{{\bf k} \sigma}
\label{eqU1}
\end{eqnarray}
the Hamiltonian stays the same; however,
the state vector becomes
\begin{eqnarray}
|{\rm BCS} (\theta) \rangle=\prod_{\bf k}\left(u_{\bf k}+v_{\bf k}c^{\dagger}_{{\bf k} \uparrow}c^{\dagger}_{-{\bf k} \downarrow}
\label{eqBCStheta}
e^{ {i}{\theta}} \right)|{\rm vac} \rangle.
\end{eqnarray}
If this phase $\theta$ is a physically meaningful parameter, the global $U(1)$ invariance is broken.  The Nambu-Goldstone mode corresponds to the spatial and time variation of $\theta$.

Independent of the BCS theory, Ginzburg and Landau developed a phenomenological theory of superconductivity using a macroscopic waver function 
\begin{eqnarray}
\Psi_{GL}=n_s^{1 /2}e^{i\theta}
\label{eqGL0}
\end{eqnarray}
where $n_s$ is the number density of the charged particle with charge $q$ for the supercurrent \cite{GL}.  The current density is given by
\begin{eqnarray}
{\bf J}=-{ {n_s q^2} \over {m_q}}\left( {\bf A}^{\rm em} +{\hbar \over {2e}}\nabla \theta \right)
\label{eqGL}
\end{eqnarray}
where $m_q$ is the mass of the charged particle and ${\bf A}^{\rm em}$ is the ordinary electromagnetic vector potential. 
Here, the factor ${\hbar \over {2e}}$ added to $\nabla \theta$ so that the observed flux quantum ${ h \over {2e}}$ is obtained.

Then, ${\bf A}$ in Eq.~(\ref{eqLondon0})
is identified as
\begin{eqnarray}
{\bf A}={\bf A}^{\rm em} +{\hbar \over {2e}}\nabla \theta
\label{eqA}
\end{eqnarray}

Comparison with Eq.~(\ref{eqLondon01}) and Eq.~(\ref{eqGL}) yields
\begin{eqnarray}
\nabla \chi_s= { \hbar \over 2 }\nabla \theta
\end{eqnarray}
Thus,  identifying the origin of  $\nabla \theta$ amounts to identifying the origin of $\nabla \chi_s$.

One origin of $\theta$ is to identify it to the $\theta$ in the BCS wave function \cite{Gorkov}. 
Gor'kov attributed $\Psi_{GL}$  to 
\begin{eqnarray}
\Psi_{GL}({\bf r})=\langle \hat{\Psi}_{\downarrow}({\bf r}) \hat{\Psi}_{\uparrow}({\bf r}) \rangle
\label{psi-s1}
\end{eqnarray}
where $\langle \hat{O} \rangle$ denotes the expectation value of the operator $\hat{O}$, and $\hat{\Psi}_{\sigma}({\bf r})$ the field operator for electrons with spin $\sigma$\cite{Gorkov} given by
\begin{eqnarray}
\hat{\Psi}_{\sigma}({\bf r})={1 \over \sqrt{\cal V}}\sum_{\bf k} e^{i {\bf k}\cdot {\bf r}}c_{{\bf k} \sigma}
\label{eqField}
\end{eqnarray}
 and ${\cal V}$ is the volume of the system. 
  
 By taking the expectation value using $|{\rm BCS} (\theta) \rangle$, we have
 \begin{eqnarray}
\Psi_{GL}={ 1 \over \cal{V}}\sum_{\bf k} u_k v_k e^{i\theta} ={ 1 \over {g {\cal V}}}\Delta_{\rm BCS}e^{i\theta}
\end{eqnarray}
where
 \begin{eqnarray}
\Delta_{\rm BCS}=g \sum_{\bf k} u_k v_k 
\end{eqnarray}
is the energy gap created by the electron-pair formation.
Gor'kov allowed a spatial variation of $\theta$; then, $\theta$ becomes a function of ${\bf r}$, $\theta({\bf r})$.

For the gauge transformation corresponding to Eq.~(\ref{eqFixchi}),
${\theta \over 2}$ transforms as
\begin{eqnarray}
{\theta \over 2} \rightarrow  {\theta \over 2}- {e \over \hbar} f
\end{eqnarray}
according to Eqs.~(\ref{eqWave}) and (\ref{eqU1}) using $q=-e$.
Then, ${\bf A}$ in Eq.~(\ref{eqA}) is invariant. Therefore, the gauge invariance of the induced current is achieved by the gauge invariance of ${\bf A}$.

The way gauge invariance of ${\bf J}$ is achieved here is different from the way it is achieved by Eq.~(\ref{eqKernel}).
In the latter, the property of the kernel $K_{\mu \nu}(q)$ is used; however, in the former, the appearance of $\theta$ is used.
We will used the appearance of $\theta$ as to the means to achieve the gauge invariance of ${\bf J}$.

The Gor'kov's identification of $\theta$ relies on the breakdown of the global $U(1)$ gauge invariance.
We will look for the origin of $\theta$ that does not require the breakdown of the global $U(1)$ gauge invariance.
 
In the following, we use the ``fictitious'' vector potential
\begin{eqnarray}
{\bf A}^{\rm fic}={\hbar \over {2e}}\nabla \theta={1 \over {e}}\nabla \chi_s
\label{eqAfic}
\end{eqnarray}
and identify the gauge invariant ${\bf A}$ as the effective gauge potential ${\bf A}^{\rm eff}$ given by
\begin{eqnarray}
{\bf A}^{\rm eff}={\bf A}^{\rm em} +{\bf A}^{\rm fic}
\label{eqAeff}
\end{eqnarray}
This ${\bf A}^{\rm eff}$ is the effective vector potential in materials, and the appearance of the nontrivial ${\bf A}^{\rm fic}$ is the key to the supercurrent generation.

\section{Reversible superconducting-normal phase transition in a magnetic field}

Reversible superconducting-normal phase transition in a magnetic field indicates that supercurrent generation mechanism needs to explain how the kinetic energy is transferable to the magnetic field energy without dissipation, and vice versa \cite{koizumi2020b}. 
It has been argued that this transformation becomes possible if the supercurrent is a collection of topologically-protected loop currents 
generated\cite{koizumi2020b}. The topological protection arises from the appearance of the nontrivial ${\bf A}^{\rm fic}$. We explain it in this section. 

 Let us consider the superconducting-normal state transition in a magnetic field using the free energy given by a sum of the magnetic field energy
\begin{eqnarray}
F_m={ 1 \over {2 \mu_0}} \int d^3 r \ {\bf B}^2
\end{eqnarray}
where ${\bf B}$ is the magnetic field,
and the kinetic energy for the supercurrent,
\begin{eqnarray}
F_k={  m_q \over {2}} \int d^3 r \ {\bf v}_s^2 n_s 
\end{eqnarray}
where ${\bf v}_s$ is the velocity of the superconducting current \cite{Fossheim2004}.

Using one of Maxwell's equations
\begin{eqnarray}
 \nabla \times {\bf B}&=&\mu_0 {\bf J}
 \label{eqB-J}
\end{eqnarray}
and the relation 
\begin{eqnarray}
{\bf J}=qn_s{\bf v}_s
\label{eqJv}
\end{eqnarray}
$F_k$ is given by
\begin{eqnarray}
F_k={  m_q \over {q^2 n_s \mu_0^2}} \int d^3 r \ (\nabla \times {\bf B})^2
\end{eqnarray}

From the minimization condition of $F_m+F_k$ with changing ${\bf B}$, we have
\begin{eqnarray}
{\bf B}+{  m_q \over {2q^2 n_s \mu_0}}\ (\nabla \times {\bf B})=0
\label{eqLondon1}
\end{eqnarray}
This explains the Meissner effect. ${\bf B}$ in the superconductor satisfies this equation.

The change of the magnetic field energy during a time interval $0 < t <\Delta t$ is given by
\begin{eqnarray}
\Delta F_m&=&{ 1 \over  {\mu_0}}\int_0^{\Delta t} dt  \int d^3 r \ {\bf B}\cdot{ {\partial {\bf B}} \over {\partial t}}
\nonumber
\\
&=&
-\int_0^{\Delta t} dt  \int d^3 r \ {\bf J}\cdot {\bf E}
\label{Fm}
\end{eqnarray}
where ${\bf E}$ is the electric field; when going from the first line to the second, one of Maxwell's equations
\begin{eqnarray}
\nabla \times {\bf E}&=&-{ {\partial {\bf B}} \over {\partial t}}
\end{eqnarray}
and another one in Eq.~(\ref{eqB-J}) are used. 
The equation (\ref{Fm}) contains the term ${\bf J}\cdot {\bf E}$, which produces the Joule heat if ${\bf J}$ contains the ordinary current.

In the standard theory of superconductivity it is assumed that the
electron pairs flow without dissipation, but single electrons flow with dissipation. Near the transition point of the superconducting-normal phase transition, a significant number of electron pairs are broken, thus, the production of the Joule heat by the flow of broken pairs is inevitable. 
This is against the fact that the superconducting-normal phase transition in a magnetic field is a reversible process \cite{Keesom}.

Now, consider the same problem using the vector potential.
We use the London equation given by
\begin{eqnarray}
{\bf J}=-{ {n_s q^2} \over {m_q}}{\bf A}^{\rm eff}=-{ {n_s q^2} \over {m_q}}({\bf A}^{\rm em} +{\bf A}^{\rm fic})
\label{eqLondon}
\end{eqnarray}
and assume that ${\bf A}^{\rm eff}$ gives rise to ${\bf B}$ in Eq.~(\ref{eqLondon1}).

Then, we rewrite ${\bf B}$ in Eq.~(\ref{eqB-J}) as ${\bf B}^{\rm eff}$ given by
\begin{eqnarray}
{\bf B}^{\rm eff}={\bf B}^{\rm em} +{\bf B}^{\rm fic}
\end{eqnarray}
where ${\bf B}^{\rm em}$ and ${\bf B}^{\rm fic}$ are defined as
\begin{eqnarray}
{\bf B}^{\rm em}&=&\nabla\times {\bf A}^{\rm em}
\\
 {\bf B}^{\rm fic}&=&\nabla\times {\bf A}^{\rm fic}
 \label{eqBfic}
\end{eqnarray}

Using the relation in Eq.~(\ref{eqJv}), 
the change of the kinetic energy due to the change of ${\bf v}_s$ during a time interval $0 < t <\Delta t$ is given by
\begin{eqnarray}
\Delta F_k&=&m_q \int_0^{\Delta t} dt   \int d^3 r \ n_s {\bf v}_s \cdot  { {\partial {\bf v}_s} \over {\partial t} }
\nonumber
\\
&=&-\int_0^{\Delta t} dt   \int d^3 r \  {\bf J} \cdot  { {\partial } \over {\partial t} }({\bf A}^{\rm eff} +{\bf A}^{\rm fic})
\end{eqnarray}

The reversible process implies $\Delta F_m+\Delta F_k=0$. 
Thus, we need to explain the following relation
\begin{eqnarray}
0= \int_0^{\Delta t} dt   \int d^3 r \  {\bf J} \cdot \left[  {\bf E}+ { {\partial} \over {\partial t} }({\bf A}^{\rm em} +{\bf A}^{\rm fic}) \right]
\label{eq7}
\end{eqnarray}

We consider the condition 
\begin{eqnarray}
 {\bf E}=-{ {\partial} \over {\partial t} }({\bf A}^{\rm em} +{\bf A}^{\rm fic})
 \label{eq10}
\end{eqnarray}
If it is satisfied, we have Eq.~(\ref{eq7}).

We consider the case where only the time-variation of ${\bf A}^{\rm fic}={ \hbar \over {2e}} \nabla \theta$ occurs.
Then, ${\bf E}$ here is ${\bf E}^{\rm fic}$ given by
\begin{eqnarray}
 {\bf E}^{\rm fic}=-{ {\partial {\bf A}^{\rm fic}} \over {\partial t} }
 \label{eqEfic}
 \end{eqnarray}
 
 Due to the multi-valuedness of $\theta$, topological quantum numbers (Chern numbers) are associated with $\theta$, and gives rise to quantized fluxoids $\Phi_k$ in Eq.~(\ref{eqFluxoid}).
We consider an abrupt change of this topological quantum number. It leads to an abrupt discrete change of ${\bf A}^{\rm fic}$, corresponding to 
the abrupt discrete change of $\Phi_k$.

The continuous change of ${\bf A}^{\rm fic}$ leads to entropy production due to the fact that the final state can be any of highly degenerate states. However, the discrete transition between topologically distinct states is a specified one characterized by the change of the whole numbers, thus, does not produce entropy. If only the latter type of transitions occur, the reversible transfer of the energy between the kinetic part and the magnetic part becomes possible.

Assuming only the time-variation from ${\bf A}^{\rm fic}$, we have
\begin{eqnarray}
\Delta F_m&=&{ 1 \over  {\mu_0}}\int_0^{\Delta t} dt  \int d^3 r \ {\bf B}^{\rm eff}\cdot{ {\partial {\bf B}^{\rm fic}} \over {\partial t}}
\nonumber
\\
&=&{ 1 \over  {\mu_0}}\int_0^{\Delta t} dt  \int d^3 r \ {\bf B}^{\rm eff}\cdot { {\partial \nabla \times {\bf A}^{\rm fic}} \over {\partial t}}
\nonumber
\\
&=&-{ 1 \over  {\mu_0}}\int_0^{\Delta t} dt  \int d^3 r \ \nabla \times {\bf B}^{\rm eff}\cdot { {\partial {\bf A}^{\rm fic}} \over {\partial t}}
\end{eqnarray}
and
\begin{eqnarray}
\Delta F_k&=&-\int_0^{\Delta t} dt   \int d^3 r \  {\bf J} \cdot  { {\partial } \over {\partial t} }{\bf A}^{\rm fic}
\nonumber
\\
&=&{ 1 \over  {\mu_0}}\int_0^{\Delta t} dt   \int d^3 r \  \nabla \times {\bf B}^{\rm eff}\cdot  { {\partial {\bf A}^{\rm fic}} \over {\partial t}}
\end{eqnarray}
thus, $\Delta F_m+\Delta F_k=0$ is established, 
where Eqs.~(\ref{eqB-J}) and (\ref{eqBfic}) are used.  

The change of ${\bf A}^{\rm fic}$ means the modification of ${\bf J}$ according to Eq.~(\ref{eqLondon}). It is noteworthy that 
the current distribution is uniquely determined by $\Phi_k$ and the boundary magnetic field as mentioned in Section~\ref{Section2}. Then, the conservation of charge induces the modification of $n_s$. The modification of ${\bf J}$ also induces the modification of ${\bf B}^{\rm eff}$ according to Eq.~(\ref{eqB-J}). Those modifications that proceed from the change of ${\bf A}^{\rm fic}$ can occur without Joule heating; thus, the
reversible superconductor-normal phase transition in a magnetic field is explained.

 \section{Berry connection for many-body wave functions and ${\bf A}^{\rm fic}$ and the number changing operators  $e^{\pm {i \over 2} \hat{\theta}({\bf r})}$}

In this section, we identify ${\bf A}^{\rm fic}$ as the {\em Berry connection for the many-body wave functions}, ${\bf A}^{\rm MB}$\cite{koizumi2019}.
 
Let us consider the wave function of a system with $N$ electrons,
\begin{eqnarray}
\Psi ({\bf x}_1, \cdots, {\bf x}_{N},t)
\label{wavef}
\end{eqnarray}
where ${\bf x}_j=({\bf r}_j,s_j) $ denotes the coordinate ${\bf r}_j$  and spin $s_j$ of the $j$th electron.

The Berry connection\cite{Berry} associated with this wave function is called the ``{\em Berry connection for many-body wave function}'' \cite{koizumi2019}. 
In order to calculate this Berry connection, we first prepare the parameterized wave function $|n_{\Psi}({\bf r}) \rangle$ with the parameter ${\bf r}$, 
 \begin{eqnarray}
\langle s, {\bf x}_{2}, \cdots, {\bf x}_{N} |n_{\Psi}({\bf r},t) \rangle = { {\Psi({\bf r}s, {\bf x}_{2}, \cdots, {\bf x}_{N},t)} \over {|C({\bf r} ,t)|^{{1 \over 2}}}}
\end{eqnarray}
where $|C({\bf r} ,t)|$ is the normalization constant given by 
\begin{eqnarray}
|C({\bf r} ,t)|=\int ds d{\bf x}_{2} \cdots d{\bf r}_{N}\Psi({\bf r} s, {\bf x}_{2}, \cdots)\Psi^{\ast}({\bf x} s, {\bf x}_{2}, \cdots)
\end{eqnarray}
where the integration with respect to $s$ actually means taking the inner product for spin functions.

Using $|n_{\Psi}\rangle$, ${\bf A}^{\rm MB}$ is given by
 \begin{eqnarray}
{\bf A}^{\rm MB}({\bf r},t)=-i \langle n_{\Psi}({\bf r},t) |\nabla_{\bf r}  |n_{\Psi}({\bf r},t) \rangle
\end{eqnarray}
Here, ${\bf r}$ is regarded as the parameter \cite{Berry}. 

When the origin of ${\bf A}^{\rm MB}$ is not the ordinary magnetic field one, i.e.,  
\begin{eqnarray}
\nabla \times {\bf A}^{\rm MB}=0
\label{BMB}
\end{eqnarray}
 it can be written in the pure gauge form,
\begin{eqnarray}
 {\bf A}^{\rm MB}=\nabla { \theta \over 2}={ e \over \hbar} {\bf A}^{\rm fic}
 \label{eqAMB}
\end{eqnarray}
where $\theta$ is a function which may be multi-valued. Actually the multi-valuedness of $\theta$ requires the existence of points where
the condition in Eq.~(\ref{BMB}) is violated. We assume that the amplitude of the wave function is zero at those singular points.

Using the relation in Eq.~(\ref{eqAMB}), $\theta$ is defined without the breakdown of the global $U(1)$ gauge invariance. 
We adopt this identification in the new theory.

The kinetic energy part of the Hamiltonian is given by
\begin{eqnarray}
K_0={ 1\over {2m}} \sum_{j=1}^{N} \left( {\hbar \over i} \nabla_{j} \right)^2
\label{a2}
\end{eqnarray}
where $m$ is the particle mass and $\nabla_{j}$ is the gradient operator with respect to the $j$th electron coordinate ${\bf r}_j$.

Using $\Psi$ and ${\bf A}^{\rm MB}$, we can construct a currentless wave function $\Psi_0$ for the current operator associated with $K_0$
\begin{eqnarray}
\Psi_0 ({\bf x}_1, \cdots, {\bf x}_{N},t)=\Psi ({\bf x}_1, \cdots, {\bf x}_{N},t)\exp\left(- i \sum_{j=1}^{N} \int_{0}^{{\bf r}_j} {\bf A}^{\rm MB}({\bf r}',t) \cdot d{\bf r}' \right)
\label{wavef0}
\end{eqnarray}

Reversely, $\Psi ({\bf x}_1, \cdots, {\bf x}_{N},t)$ is expressed as
 \begin{eqnarray}
\Psi =\Psi_0\exp\left({ i  \over 2} \sum_{j=1}^{N} \theta ({\bf r}_j, t) \right)
\label{f}
\end{eqnarray}
using the currentless wave function $\Psi_0$.

Let us obtain the conjugate momentum of $\theta$. 
 For this purpose, we use the time-dependent variational principle using the following Lagrangian \cite{Koonin1976},
\begin{eqnarray}
{\cal L}\!=\langle \Psi | i\hbar \partial_t \!-\!H| \Psi \rangle\!=\! i\hbar \langle \Psi_0 | \partial_t | \Psi_0 \rangle- {\hbar \over 2} \int \!d{\bf r} \ {{n \dot{\theta} }} - \langle \Psi |H| \Psi \rangle
\label{L}
\end{eqnarray}
where $H$ is the Hamiltonian and $n$ is the number density of the particles.

From the above Lagrangian, the conjugate momentum of $\theta$ is obtained as
\begin{eqnarray}
p_{\theta}= {{\delta {\cal L}} \over {\delta \dot{\theta}}}=-{\hbar \over 2} n
\label{momentumchi}
\end{eqnarray}
thus, $\theta$ and $n$ are canonical conjugate variables.

If we follow the canonical quantization condition 
\begin{eqnarray}
[\hat{p}_{\theta}({\bf r}, t), \hat{\theta}({\bf r}', t)]=-i\hbar \delta ({\bf r}- {\bf r}')
\end{eqnarray}
where $\hat{p}_{\theta}$ and $\hat{\theta}$ are operators corresponding to ${p}_{\theta}$ and ${\theta}$ respectively, 
we have
\begin{eqnarray}
\left[{ {\hat{n}({\bf r}, t)} } , { 1 \over 2} \hat{\theta}({\bf r}', t) \right]=i \delta ({\bf r}- {\bf r}')
\label{commu0}
\end{eqnarray}
where $\hat{n}$ is the operator corresponding to $n$.
 Strictly speaking, $\hat{\theta}$ is not a hermitian operator; however, it is known that when it is used as 
 $\sin \hat{\theta}$ or  $\cos \hat{\theta}$, the problem is avoided, practically \cite{Phase-Angle}. In the following we use
 $\hat{\theta}$  as $e^{\pm i {\hat{\theta} }}$.
 
We construct the following boson field operators from $\hat{\theta}$ and $\hat{n}$,
\begin{eqnarray}
\hat{\psi}^{\dagger}({\bf r})= \left(\hat{n}({\bf r}) \right)^{1/2} e^{- {i \over 2}{\hat{\theta}({\bf r}) }}, \quad \hat{\psi}({\bf r})= e^{{i \over 2} { \hat{\theta}({\bf r}) }}\left( \hat{n}({\bf r}) \right)^{1/2}
\label{boson1}
\end{eqnarray}

Using Eq.~(\ref{commu0}), the following relations are obtained,
\begin{eqnarray}
 [\hat{\psi}({\bf r}),\hat{\psi}^{\dagger}({\bf r}')]=\delta({\bf r}-{\bf r}'), \quad  [{\psi}({\bf r}),\hat{\psi}({\bf r}')]=0, \quad  [\hat{\psi}^{\dagger}({\bf r}),\hat{\psi}^{\dagger}({\bf r}')]=0
 \label{commu2}
\end{eqnarray}
   
 The number operator for the particle participating in the collective mode described by $\theta$ is given by
 \begin{eqnarray}
\hat{N}_{\theta}=\int_{\cal V } d{\bf r}  \hat{\psi}^{\dagger}({\bf r}) \hat{\psi}^{}({\bf r}) 
 \label{commu3}
 \end{eqnarray}
 
 The following relation is obtained from Eqs.~(\ref{boson1}), (\ref{commu2}), and (\ref{commu3}),
   \begin{eqnarray}
[e^{ {i \over 2} \hat{\theta}({\bf r})}, \hat{N}_{\theta}]=e^{ { i \over 2} \hat{\theta}({\bf r})}
\label{commchi}
 \end{eqnarray}
 
We define eigenstates of $\hat{N}_{\theta}$ as follows, 
    \begin{eqnarray}
\hat{N}_{\theta} | N_{\theta}\rangle =N_{\theta}| N_{\theta} \rangle
 \end{eqnarray} 

 Then, $e^{ { \pm {i \over 2} } \hat{\theta}}$ are the number changing operators that satisfy
    \begin{eqnarray}
  e^{\pm {i \over 2} \hat{\theta}({\bf r})} | N_{\theta} \rangle = e^{\pm {i \over 2} {\theta}({\bf r})} | N_{\theta} \mp 1 \rangle
\label{commchi2}
 \end{eqnarray}
where the phase on $| N_{\theta} \mp 1 \rangle$ is so chosen that it incorporates the phase factor from the Berry connection \cite{koizumi2019}.
This indicates that $e^{\pm {i \over 2}\hat{\theta}({\bf r})}$ are the number changing operators for the number of particles participating in the collective mode described by $\theta$, and it provides the phase that arises from the Berry connection in the state $| N_{\theta} \rangle$.

\section{The particle number conserving Bogoliubov transformation and the particle number conserving  Bogoliubov-de Gennes equations}

From the BCS model, a relation corresponding to Eq.~(\ref{commu0}) is obtained as follows: first, we divide the system into coarse-grained cells of unit volumes; and express the BCS state in the coarse-grained cell with the center position ${\bf r}$ as
\begin{eqnarray}
|\Psi_{\rm BCS} ({\bf r},t) \rangle=\prod_{\bf k} \left( u_{k} ({\bf r},t) + e^{i \theta({\bf r},t) } v_{k} ({\bf r},t) c^{\dagger}_{{\bf k} \uparrow}c^{\dagger}_{-{\bf k} \downarrow} \right) |{\rm vac} \rangle
\label{BCSr}
\end{eqnarray}
Now $u_k$, $v_k$, and $\theta$ depend on the coordinate and time. 

Then, the Lagrangian corresponding to Eq.~(\ref{L}) is given by
\begin{eqnarray}
{\cal L}_{\rm BCS}\!&=&\int d^3r\langle \Psi_{\rm BCS} ({\bf r},t)  | i\hbar \partial_t \!-\!H_{\rm BCS}| \Psi_{\rm BCS} ({\bf r},t)  \rangle\!
\nonumber
\\
&=&-\!\int \!d^3r \ {{n_{e} \dot{\theta} \hbar} \over 2}\!-\! \int d^3r \langle \Psi_{\rm BCS}({\bf r},t)  | H_{\rm BCS} | \Psi_{\rm BCS} ({\bf r},t)  \rangle\!
\label{L2}
\end{eqnarray}
where $H_{\rm BCS}$ is now coordinate dependent and specified by the coarse-grained cell position ${\bf r}$.
The electron number density in the cell is given by
\begin{eqnarray}
n_{e}({\bf r},t)= 2\sum_{\bf k} v^2_{k} ({\bf r},t)
\label{rhoBCS}
\end{eqnarray}

If we identify $n_{e}({\bf r},t)$ as $n({\bf r},t)$ in Eq.~(\ref{L}), the number changing operators $e^{\pm{i \over 2} \hat{\theta}({\bf r},t) }$ are defined, analogously. 

It is well-known that $|\Psi_{\rm BCS} ({\bf r},t) \rangle$ can be considered as the vacuum for the Bogoliubov type excitations; 
if we define the Bogoliubov operators, 
\begin{eqnarray}
\gamma_{{\bf k} 0}&=&u_{k}  e^{ -{i \over2}{\theta}}c_{{\bf k} \uparrow}- v_{k}  c^{\dagger}_{-{\bf k} \downarrow} e^{ {i \over2}{\theta}}
\nonumber
\\
\gamma_{-{\bf k} 1} &=& u_{k} e^{- {i \over2}{\theta}}c_{-{\bf k} \downarrow}+ v_{k}  c^{\dagger}_{{\bf k} \uparrow} e^{ {i \over2}{\theta}}
\label{eqB}
\end{eqnarray}
they satisfy 
\begin{eqnarray}
&&\gamma_{{\bf k} 0}|\Psi_{\rm BCS} ({\bf r},t) \rangle=0, \quad \gamma_{-{\bf k} 1} |\Psi_{\rm BCS} ({\bf r},t) \rangle=0
\nonumber
\\
&&\langle \Psi_{\rm BCS} ({\bf r},t) | \gamma^{\dagger}_{{\bf k} 0}=0, \quad \langle \Psi_{\rm BCS} ({\bf r},t) |\gamma^{\dagger}_{-{\bf k} 1}=0
\end{eqnarray}

In the following, we regard the superconducting state as the vacuum of Bogoliubov type excitations. However, we use the particle number conserving version of the Bogoliubov operators; 
we replace $e^{\pm i{ 1 \over 2}{\theta}({\bf r})}$ in Eq.~(\ref{eqB})
to $e^{\pm i{ 1 \over 2}\hat{\theta}({\bf r})}$, where
$e^{\pm i{ 1 \over 2}\hat{\theta}({\bf r})}$ satisfy
\begin{eqnarray}
  e^{\pm {i \over 2} \hat{\theta}({\bf r})} | {\rm Gnd}(N)\rangle = e^{\pm {i \over 2} {\theta}({\bf r})} | {\rm Gnd}(N \mp 1) \rangle
  \label{Gnd}
 \end{eqnarray}
 by following Eq.~(\ref{commchi2}), where $| {\rm Gnd}(N)\rangle$ is the superconducting ground state with $N$ electrons.

We denote the particle number conserving Bogoliubov operators by $\gamma_{n \sigma}$ and $\gamma^{\dagger}_{n \sigma}$; they satisfy 

\begin{eqnarray}
\gamma_{n \sigma}|{\rm Gnd}(N) \rangle=0; \quad  \langle {\rm Gnd}(N)| \gamma^{\dagger}_{n \sigma}=0
\end{eqnarray}
 
We consider the Bogoliubov-de Gennes formalism using $\gamma_{n \sigma}$ and $\gamma^{\dagger}_{n \sigma}$.
  The electron field operators, which are given previously in Eq.~(\ref{eqField}), are now written as 
 \begin{eqnarray}
\hat{\Psi}_{\uparrow}({\bf r}) &=&\sum_n e^{i{ 1 \over 2}\hat{\theta}({\bf r})} ( \gamma_{n \uparrow}u_n({\bf r}) - \gamma^{\dagger}_{n \downarrow}v^{\ast}_n({\bf r}) 
)
\nonumber
\\
\hat{\Psi}_{\downarrow}({\bf r}) &=&\sum_n e^{i{ 1 \over 2}\hat{\theta}({\bf r})} ( \gamma_{n \downarrow}u_n({\bf r}) + \gamma^{\dagger}_{n \uparrow}v^{\ast}_n({\bf r}) 
)
\label{eqFiledOp}
\end{eqnarray}
Note that $e^{i{ 1 \over 2}\hat{\theta}({\bf r})}$ reduces the particle number by one, and $\gamma_{n \sigma}$ and $\gamma^{\dagger}_{n \sigma}$ do not change the particle number, thus, $\hat{\Psi}_{\sigma}({\bf r})$ reduces the particle number by one.

The effective Hamiltonian of the Bogoliubov-de Gennes formalism  is given by 
\begin{eqnarray}
{\cal H}_{\rm eff}&=&\int d{\bf r} \Big[
\sum_{\sigma} \left( \hat{\Psi}^{\dagger}_{\sigma}({\bf r}) {\cal H}_e \hat{\Psi}_{\sigma}({\bf r})
+U({\bf r})  \hat{\Psi}^{\dagger}_{\sigma}({\bf r}) \hat{\Psi}_{\sigma}({\bf r}) \right) 
\nonumber
\\
&+&\Delta({\bf r})e^{i\hat{\theta}({\bf r})} \hat{\Psi}^{\dagger}_{\uparrow}({\bf r}) \hat{\Psi}^{\dagger}_{\downarrow}({\bf r}) + \Delta^{\ast}({\bf r})e^{-i\hat{\theta}({\bf r})} \hat{\Psi}_{\downarrow}({\bf r}) \hat{\Psi}_{\uparrow}({\bf r}) 
\Big]
\label{Heff}
\end{eqnarray}
where
  ${\cal H}_e$, $U({\bf r})$, and $\Delta({\bf r})$ are given by
\begin{eqnarray}
{\cal H}_e &=& { 1 \over {2m}} 
\left( -i\hbar \nabla -q{\bf A}^{\rm em} \right)^2 +U_0({\bf r})-{\cal E}_F
\label{eqHe}
\\
U({\bf r})&=&-g\langle   \hat{\Psi}^{\dagger}_{\uparrow }({\bf r}) \hat{\Psi}_{\uparrow}({\bf r}) \rangle 
=- g\langle   \hat{\Psi}^{\dagger}_{\downarrow }({\bf r}) \hat{\Psi}_{\downarrow}({\bf r}) \rangle 
\\
\Delta({\bf r})&=&-g \langle   e^{-i \hat{\theta}({\bf r})}\hat{\Psi}_{\downarrow }({\bf r}) \hat{\Psi}_{\uparrow}({\bf r}) \rangle 
=g \langle e^{-i \hat{\theta}({\bf r})} \hat{\Psi}_{\uparrow }({\bf r}) \hat{\Psi}_{\downarrow}({\bf r}) \rangle 
\end{eqnarray}

The particle number conserving Bogoliubov operators $\gamma_{n \sigma}$ and $\gamma^{\dagger}_{n \sigma}$ obey fermion commutation relations. They are chosen to satisfy
\begin{eqnarray}
\left[ {\cal H}_{\rm eff}, \gamma_{n \sigma } \right] &=&-\epsilon_n \gamma_{n \sigma}
\\
\left[{\cal H}_{\rm eff}, \gamma^{\dagger}_{n \sigma } \right] &=&\epsilon_n \gamma^{\dagger}_{n \sigma}
\end{eqnarray}
with $\epsilon_n \geq 0$. Then, ${\cal H}_{\rm eff}$ is diagonalized as
\begin{eqnarray}
{\cal H}_{\rm eff}=E_g + \sum_{n, \sigma} \epsilon_n \gamma^{\dagger}_{n \sigma}\gamma_{n \sigma}
\end{eqnarray}
where $E_g$ is the ground state energy.

Using Eq.~(\ref{Heff}) and commutation relations for $\hat{\Psi}^{\dagger}_{\sigma }({\bf r})$ and $\hat{\Psi}_{\sigma }({\bf r})$,
\begin{eqnarray}
&&\{ \hat{\Psi}_{\sigma }({\bf r}),\hat{\Psi}^{\dagger}_{\sigma' }({\bf r}') \}=\delta_{\sigma \sigma'}\delta({\bf r} -{\bf r}')
\nonumber
\\
&&\{ \hat{\Psi}_{\sigma }({\bf r}),\hat{\Psi}_{\sigma' }({\bf r}') \}=0
\nonumber
\\
&&\{ \hat{\Psi}^{\dagger}_{\sigma }({\bf r}),\hat{\Psi}^{\dagger}_{\sigma' }({\bf r}') \}=0
 \end{eqnarray}
the following relations are obtained
\begin{eqnarray}
\left[\hat{\Psi}_{\uparrow }({\bf r}) , {\cal H}_{\rm eff} \right] &=&
\left[{\cal H}_e + U({\bf r}) \right] \hat{\Psi}_{\uparrow }({\bf r})+\Delta({\bf r}) e^{i \hat{\theta}({\bf r})}\hat{\Psi}^{\dagger}_{\downarrow }({\bf r})
\label{deG1}
\\
\left[\hat{\Psi}_{\downarrow }({\bf r}) , {\cal H}_{\rm eff} \right] &=&
\left[{\cal H}_e + U({\bf r}) \right] \hat{\Psi}_{\downarrow }({\bf r})-\Delta({\bf r}) e^{i \hat{\theta}({\bf r})}\hat{\Psi}^{\dagger}_{\uparrow }({\bf r})
\label{deG2}
\end{eqnarray}

By taking into account the relation in Eq.~(\ref{Gnd}), we can replace $e^{i{ 1 \over 2}\hat{\theta}({\bf r})}$ with $e^{i{ 1 \over 2}{\theta}({\bf r})}$. 

Finally, we obtain the following system of equations,
\begin{eqnarray}
\epsilon_n u_n({\bf r})&=&
\left[\bar{\cal H}_e + U({\bf r}) \right] u_n({\bf r})+\Delta ({\bf r})v_n({\bf r})
\nonumber
\\
\epsilon_n v_n({\bf r})&=&-
\left[\bar{\cal H}^{\ast}_e + U({\bf r}) \right] v_n({\bf r})+\Delta^{\ast}({\bf r})u_n({\bf r})
\label{eq4-27}
\end{eqnarray}
where
\begin{eqnarray}
U({\bf r})&=&-g\sum_n |v_n({\bf r})|^2
\\
\Delta({\bf r})&=&g\sum_n v^{\ast}_n({\bf r})  u_n({\bf r})
\\
\bar{\cal H}_e &=& { 1 \over {2m}} 
\left( -i\hbar \nabla -q{\bf A}^{\rm eff} \right)^2 +U_0({\bf r})-E_F
\end{eqnarray}
This is the system of equations for $\epsilon_n,  v_n({\bf r})$, $u_n({\bf r})$.
Note that the gauge invariant ${\bf A}^{\rm eff}$ appears instead of ${\bf A}^{\rm em}$. 

From the field operators in Eq.~(\ref{eqFiledOp}), the annihilation and creation operators for electrons with spin $\sigma$ at the $i$th site, $c_{ i \sigma}$ and $ c^{\dagger}_{ i \sigma}$, 
are obtained as
  \begin{eqnarray}
 c_{ i \sigma} &=&\sum_{n}[ u^{n}_{i}\gamma_{n \sigma}-\sigma (v^{n}_{i})^{\ast}\gamma_{n -\sigma}^{\dagger}] e^{ {i \over 2} \hat{\theta}_i}
 \nonumber
 \\
 c^{\dagger}_{ i \sigma} &=&\sum_{n}[ (u^{n}_{i \sigma})^{\ast}\gamma^{\dagger}_{n \sigma}-\sigma v^{n}_{i}\gamma_{n -\sigma}] e^{-{i \over 2} \hat{\theta}_i}
 \label{Bog}
  \end{eqnarray}
 where $\sigma$ is the spin, $+1$ for $\uparrow$, and $-1$ for $\downarrow$. These relations are used to calculate physical quantities in the superconducting state which is described by $e^{ {i \over 2} \hat{\theta}_i}$,  $e^{ -{i \over 2} \hat{\theta}_i}$, $\gamma_{n \sigma}$, and $\gamma^{\dagger}_{n \sigma}$.
 
For the gauge transformation 
\begin{eqnarray}
{\bf A}^{\rm eff} \rightarrow {\bf A}^{\rm em}
\end{eqnarray}
 $u_n$ and $v_n$ transform as
\begin{eqnarray}
u_n &\rightarrow&  u_n e^{-{ i \over \hbar} q \int^{\bf r} d{\bf r}' \cdot {\bf A}^{\rm fic}({\bf r}')}=u_n e^{{ i \over 2} \theta}
\nonumber
\\
v_n &\rightarrow&  v_n e^{{ i \over \hbar} q \int^{\bf r} d{\bf r}' \cdot {\bf A}^{\rm fic}({\bf r}')}=v_n e^{-{ i \over 2} \theta}
\end{eqnarray}
according to Eq.~(\ref{eq4-27}), 

Thus, $\Delta({\bf r})$ transforms as
\begin{eqnarray}
\Delta({\bf r}) \rightarrow&  \Delta({\bf r})e^{ i \theta}
\end{eqnarray}
It has the same $\theta$ dependence as $\Psi_{GL}$ in Eq.~(\ref{eqGL0}). Thus, we may use $\Delta({\bf r})$ for $\Psi_{GL}$ if $\Delta({\bf r})\neq0$ coincides with the appearance of the superconducting state. We believe this corresponds to the case where the standard theory works very well.

\section{Josephson Tunneling using the particle number conserving Bogoliubov operators}
In this section, we investigate Josephson tunneling using the particle number conserving version of Bogoliubov operators.
 
We use the following electron transfer Hamiltonian for the Josephson junction, 
  \begin{eqnarray}
H_{LR}=-\sum_{\sigma}T_{LR} \left(  c^{\dagger}_{L \sigma} c_{R \sigma}+c^{\dagger}_{R\sigma} c_{L \sigma}  \right)
\label{juncH}
\end{eqnarray}
Labels``$L$'' and ``$R$'' refer to the left superconductor $S_L$ and right superconductor $S_R$ in Fig.~\ref{Josephson}, respectively.

Using the Boboliubov transformation in Eq.~(\ref{Bog}) and including the electromagnetic field by the Peierls substitution, $H_{LR}$ is rewritten as
  \begin{eqnarray}
H_{LR}&=&-T_{L R} e^{- { i \over 2}(\hat{\theta}_L-\hat{\theta}_R)} e^{-i {e \over \hbar} \int_R^L d{\bf r} \cdot {\bf A}^{\rm em}}
\nonumber
\\
&& \times
\sum_{n,m} \Big[
(
(u^{n}_{L})^{\ast}\gamma^{\dagger }_{n \downarrow} + v^{n}_{L }\gamma_{n \uparrow}) ( u^{m}_{R }\gamma_{m \downarrow}+ (v^{m}_{R})^{\ast}\gamma_{m \uparrow}^{\dagger} ) 
\nonumber
\\
&&+
(
(u^{n}_{L })^{\ast}\gamma^{\dagger }_{n \uparrow} - v^{n}_{L}\gamma_{n \downarrow}) ( u^{m}_{R }\gamma_{m \uparrow}- (v^{m}_{R})^{\ast}\gamma_{m  \downarrow}^{\dagger} )  
\Big]+\mbox{h.c.}
\end{eqnarray}
where we have assumed that the two superconductors in the junction have a common set of Bogoliubov operators. The use of the common operators means that the two superconductors is a united one. This possibility was not considered in the Josephson's derivation \cite{Josephson62}.

Taking expectation value of $H_{LR}$ with respect to the junction ground state, we obtain the junction energy
\begin{eqnarray}
H_J^{e}=C \cos \left[{ {e \over \hbar} \int_R^L d{\bf r} \cdot \left({\bf A}^{\rm em} +{\hbar \over {2e}} \nabla \theta\right)} + \alpha \right]
\label{e}
\end{eqnarray}
where
\begin{eqnarray}
{ 1 \over 2} C e^{i \alpha}=-2T_{L R} \sum_{n}
 v^{n}_{L }(v^{n}_{R})^{\ast}+\mbox{h.c.}
\end{eqnarray}

If we assume that the two superconductors in the junction have different sets of Bogoliubov operators as considered in the Josephson's derivation \cite{Josephson62},  $H_{LR}$ is rewritten as
  \begin{eqnarray}
H_{LR}&=&-T_{L R} e^{- { i \over 2}(\hat{\theta}_L-\hat{\theta}_R)} e^{-i {e \over \hbar} \int_R^L d{\bf r} \cdot {\bf A}^{\rm em}}
\nonumber
\\
&& \times
\sum_{n,m} \Big[
(
(u^{n}_{L})^{\ast}\gamma^{\dagger }_{L n \downarrow} + v^{n}_{L }\gamma_{L n \uparrow}) ( u^{m}_{R }\gamma_{R m \downarrow}+ (v^{m}_{R})^{\ast}\gamma_{R m \uparrow}^{\dagger} ) 
\nonumber
\\
&&+
(
(u^{n}_{L })^{\ast}\gamma^{\dagger }_{L n \uparrow} - v^{n}_{L}\gamma_{L n \downarrow}) ( u^{m}_{R }\gamma_{R m \uparrow}- (v^{m}_{R})^{\ast}\gamma_{R m  \downarrow}^{\dagger} )  
\Big]+\mbox{h.c.}
\end{eqnarray}
where the Bogoliubov operators have labels for the superconductors, $L$ for the left and $R$ for the right superconductors.
In this case, we need to employ the second order perturbation theory to have nonzero expectation value for the matrix element calculated with the ground state.

From the second order perturbation, the effective interaction Hamiltonian is calculated as
\begin{eqnarray}
&&\langle H_{LR}{1 \over {E_0 - H_0}} H_{LR} \rangle \approx - \langle \sum_{m, n, m', n'}T_{L R}^2 
\left[ e^{ -{i \over 2} (\hat{\theta}_L-\hat{\theta}_R)}v_L^{n}u_R^{m}(\gamma_{L n \uparrow} \gamma_{R m \downarrow}-\gamma_{L n \downarrow}\gamma_{R m \uparrow})+ (L \leftrightarrow R) 
\right]
\nonumber
\\
&\times& {1 \over {\epsilon_m^{R}+\epsilon_n^{L}}}
\left[ e^{ -{i \over 2} (\hat{\theta}_L-\hat{\theta}_R)}(u_{L}^{ n'}v_{R}^{ m'})^{\ast} (\gamma^{\dagger}_{L n' \downarrow} \gamma^{\dagger}_{R m' \uparrow}- \gamma^{\dagger}_{L n' \uparrow} \gamma^{\dagger}_{R m' \downarrow})+ (L \leftrightarrow R) 
\right] \rangle
\nonumber
\\
&\approx& - \sum_{m, n} {{2T_{L R}^2 } \over {\epsilon_m^{R}+\epsilon_n^{L}}}
\left[ v_L^{n}u_R^{m} (u_{L}^{ n}v_{R}^{ m})^{\ast}e^{- {i } (\hat{\theta}_L-\hat{\theta}_R)}+(v_L^{n}u_R^{m})^{\ast} u_{L}^{ n}v_{R}^{ m}e^{ {i} (\hat{\theta}_L-\hat{\theta}_R)})
+|u_{L}^{ n}v_{R}^{ m}|^2+ |v_{L}^{ n}u_{R}^{ m}|^2\right]
\nonumber
\\
\label{Perttransfer3}
\end{eqnarray}

Including the vector potential ${\bf A}^{\rm em}$, the junction energy for the Josephson effect is given by
\begin{eqnarray}
H_J^{2e}=C' \cos \left({ {{2e} \over \hbar} \int_R^L d{\bf r} \cdot ({\bf A}^{\rm em} +{\hbar \over {2e}} \nabla \theta)} + \alpha' \right)
\label{2e}
\end{eqnarray}
where $C'$ and $\alpha'$ are constants.
This formula gives rise to the Ambegaokar-Baratoff relation\cite{Ambegaokar} for the dc Josephson effect\cite{Josephson62}.

Whether $H_J^{e}$ or $H_J^{2e}$ is appropriate to describe the Josephson effect depends on the junction. 
In the standard theory $H_J^{2e}$ is used to have Eq.~(\ref{eqJR}).
However, it has been shown that $H_J^{e}$ is the right one to use to obtain Eq.~(\ref{eqJR}) in the real experimental boundary condition \cite{Koizumi2011,HKoizumi2015}. Let us see this point, below.

If we use $H_J^{e}$, $\dot{\phi}$ in Eq.~(\ref{eqJR}) is given by
\begin{eqnarray}
 \dot{\phi}&=&{e \over \hbar} \int_R^L d{\bf r} \cdot \left(\partial_t{\bf A}^{\rm em} +{\hbar \over {2e}} \nabla \partial_t{\theta} \right)
 \nonumber
 \\
 &=&-{e \over \hbar} \int_R^L d{\bf r} \cdot {\bf E}^{\rm em}  +\left. {e \over \hbar}  \left( - \varphi^{\rm em} +{\hbar \over {2e}}  \partial_t{\theta} \right)\right|^L_R
 \label{eqnPhidot}
 \end{eqnarray}
where the identity 
\begin{eqnarray}
 {\bf E}^{\rm em} =-\partial_t{\bf A}^{\rm em} - \nabla \varphi^{\rm em}
 \end{eqnarray}
is used.

The voltage across the junction $V$ is given by
\begin{eqnarray}
V=- \int_R^L d{\bf r} \cdot {\bf E}^{\rm em}
  \end{eqnarray}
  
 Now we consider the second term in Eq.~(\ref{eqnPhidot}).  
 \begin{eqnarray}
   \varphi^{\rm em} -{\hbar \over {2e}}  \partial_t{\theta} 
  \end{eqnarray}
 is actually the time component partner of ${\bf A}^{\rm eff}$ in Eq.~(\ref{eqAeff}).
  Thus, we write it as 
   \begin{eqnarray}
  \varphi^{\rm eff}= \varphi^{\rm em} -{\hbar \over {2e}}  \partial_t{\theta} 
  \end{eqnarray}
  It is gauge invariant like ${\bf A}^{\rm eff}$.
  
  In the Hamiltonian, it enters as a term
     \begin{eqnarray}
 \int  d^3 r \varphi^{\rm eff} \rho
   \end{eqnarray}
 thus, we relate it to the chemical potential $\mu$ 
       \begin{eqnarray}
 \mu=e \varphi^{\rm eff}
   \end{eqnarray}
   
   Then,  Eq.~(\ref{eqnPhidot}) becomes
   \begin{eqnarray}
 \dot{\phi}={e \over \hbar} V + {1 \over \hbar}  \left( \mu_R-\mu_L \right)
 \label{eqJ1}
 \end{eqnarray}

Due to the balance between the chemical potential difference and the voltage, we have
  \begin{eqnarray}
 eV =\mu_R-\mu_L 
 \end{eqnarray}
The substitution of the above in Eq.~(\ref{eqJ1}) yields the Josephson relation in Eq.~(\ref{eqJR}). Thus, we obtain the
observed Josephson relation from $H_J^{e}$. 
 
 \section{Spin-twisting itinerant motion of electrons and the appearance of nontrivial ${\bf A}^{\rm fic}$}
  \label{section8}
  
We present a possible origin of the non-trivial ${\bf A}^{\rm fic}$ in Eq.~(\ref{eqAMB}) in this section.

 We consider the modification of the kinetic energy by the following Rashba interaction term 
\begin{eqnarray}
H_{so}= {\bm \lambda}({\bf r})\cdot {{\hbar{\bm \sigma}} \over 2} \times \left[ \hat{\bf p}-{q}{\bf A}^{\rm em}({\bf r})\right], 
\end{eqnarray}
where ${\bm \lambda}({\bf r})$ is the spin-orbit coupling vector (its direction is the internal electric field direction), $\hat{\bf p}=-i\hbar \nabla$ is the momentum operator, and $q=-e$ is electron charge \cite{Rashba}. 

This term will be added in Eq.~(\ref{eqHe}), which corresponds to the Hamiltonian for the Bloch electrons when $g=0$. 

 By including the above term, the band energy for electrons with dispersion ${\cal E}({\bf k})$ is given by
\begin{eqnarray}
{\cal E}({\bf k})+\hbar{\bm \lambda}({\bf r})\times {\bf s}({\bf r})\cdot {\bf k}
\label{NewB}
\end{eqnarray}

We consider the spin-twisting itinerant motion of electrons using the coordinate depending spin function
\begin{eqnarray}
 \Sigma_1({\bf r})=e^{-{ i \over 2} \chi({\bf r})} 
 \left(
 \begin{array}{c}
 e^{-i { 1 \over 2} \xi ({\bf r})} \sin {{\zeta ({\bf r}) } \over 2}
 \\
  e^{i { 1 \over 2} \xi {\bf r})} \cos {{\zeta ({\bf r}) } \over 2} 
 \end{array}
 \right)
 \label{spin-d1}
\end{eqnarray}
where $\zeta ({\bf r})$ and $\xi ({\bf r})$ are the polar and azimuthal angles of the spin-direction, respectively.

The important point here is that the phase factor $e^{-{ i \over 2} \chi({\bf r})}$ is added to make it a single-valued function of the coordinate ${\bf r}$. Here, the orbital part of the wave function is assumed to be single-valued.
This requirement is equivalent to the assumption that the basis ${ |{\bf r} \rangle}$ composed of eigenstates of the coordinate operator $\hat{\bf r}$ with the eigenvalue ${\bf r}$,
\begin{eqnarray}
\hat{\bf r}|{\bf r} \rangle ={\bf r}|{\bf r} \rangle
\end{eqnarray}
exists. When the electrons perform spin-twisting itinerant circular motion, it may cause the sign change due to its spinor property;  however, such a state must have the single-valued representation in the coordinate representation.

We may rephrase the above argument as follows;  when the spin state is given by 
\begin{eqnarray}
| \Sigma_1 \rangle
\end{eqnarray}
 its coordinate representation is given by
\begin{eqnarray}
\langle {\bf r} | \Sigma_1 \rangle= \Sigma_1({\bf r})
\end{eqnarray}
It should be a single-valued function of ${\bf r}$ since $|{\bf r} \rangle$ is the eigenstate with a definite value of ${\bf r}$.

It appears that when the electrons perform spin-twisting itinerant circular motion, zeros of the wave function arise at the spinning centers and
a flux that gives riser to this phase factor is generated through the string of the zeros. They may be considered as the $\pi$-flux Dirac string \cite{Monopole, koizumi2020}. We will specify the condition on $\chi$ to make $\Sigma_1({\bf r})$ single-valued, later.

The expectation value of spin ${\bf s}({\bf r})=(s_x({\bf r}), s_y({\bf r}), s_z({\bf r}))$ for $\Sigma_1({\bf r})$ is calculated as
\begin{eqnarray}
s_x ({\bf r})= { \hbar \over 2} \cos \xi ({\bf r}) \sin \zeta ({\bf r}), \ s_y ({\bf r})= { \hbar \over 2} \sin \xi ({\bf r}) \sin \zeta ({\bf r}), \ s_z= { \hbar \over 2}  \cos \zeta ({\bf r})
\end{eqnarray}

The Berry connection arising from it is given by
\begin{eqnarray}
{\bf A}^{\rm fic}_1({\bf r})=-i {\hbar \over e}\Sigma_1^{\dagger} \nabla \Sigma_1= -{\hbar \over {2e}} \nabla \chi ({\bf r}) +{ \hbar \over {2e}} \nabla \xi ({\bf r}) \cos \zeta ({\bf r})
\label{eqAfic1}
\end{eqnarray}

Let us see ${\bf A}^{\rm fic}_1({\bf r})$ gives rise to a quantized cyclotron motion even without external magnetic field, below.

We quantize the cyclotron orbit by following the Onsager's argument \cite{Onsager1952}. 
Due to the presence of ${\bf A}^{\rm fic}_1({\bf r})$, the effective magnetic field ${\bf B}^{\rm eff}$ is given by
\begin{eqnarray}
{\bf B}^{\rm eff} =\nabla \times {\bf A}_1^{\rm eff}, \quad  {\bf A}_1^{\rm eff}={\bf A}^{\rm em}+{\bf A}_1^{\rm fic}
\end{eqnarray}
Thus, the Bohr-Sommerfeld relation becomes
\begin{eqnarray}
\oint_C (\hbar {\bf k} -e{\bf A}_1^{\rm eff}) \cdot d{\bf r}=2\pi \hbar \left(n+{ 1 \over 2} \right)
\label{Onsager1}
\end{eqnarray}
where $n$ is an integer and $C$ is the closed loop that corresponds to the section of Fermi surface enclosed by the cyclotron orbit.

Using the equation of motion 
\begin{eqnarray}
\hbar \dot{\bf k}=-e \dot{\bf r} \times {\bf B}^{\rm eff}
\end{eqnarray}
in the magnetic field ${\bf B}^{\rm eff}$
we have 
\begin{eqnarray}
\oint_C \hbar {\bf k} \cdot d{\bf r}&=&-e \oint_C d{\bf r} \cdot {\bf r}\times {\bf B}^{\rm eff}=e \oint_C  {\bf B}^{\rm eff} \cdot  {\bf r} \times d{\bf r}
\nonumber
\\
&=&2e \oint_C  {\bf A}_1^{\rm eff} \cdot d{\bf r}
\end{eqnarray}

Thus, Eq.~(\ref{Onsager1}) becomes
\begin{eqnarray}
e \oint_C  {\bf A}^{\rm em} \cdot  d{\bf r}+e \oint_C  {\bf A}_1^{\rm fic} \cdot  d{\bf r}=2\pi \hbar \left(n+{ 1 \over 2} \right)
\label{Onsager2}
\end{eqnarray}

The quantization condition in Eq.~(\ref{Onsager2}) is satisfied even if the magnetic field is absent. 
If the magnetic field is absent, the first term is zero. Thus, we have
\begin{eqnarray}
- \oint_C {1 \over 2} \nabla \chi ({\bf r})\cdot  d{\bf r} + \oint_C { 1 \over 2} \nabla \xi ({\bf r}) \cos \zeta ({\bf r}) \cdot  d{\bf r}=2\pi \left(n+{ 1 \over 2} \right)
\end{eqnarray}
where Eq.~(\ref{eqAfic1}) is used.

The above can be satisfied by the following two sets of conditions; one is $\zeta=\pi/2$, $w_C[\chi]=-1$, and $n=0$, and the other is $\zeta=\pi/2$, $w_C[\chi]=1$, and $n=-1$, where
\begin{eqnarray}
w_C[\chi]= {1 \over {2\pi}} \oint_C\nabla \chi ({\bf r})\cdot  d{\bf r}
\end{eqnarray}
is the winding number of $\chi$ along loop $C$. 

We will show later that the condition $\zeta=\pi/2$ is obtained as the condition for
the kinetic energy gain if the electron-pairing occurs. 

By adopting the condition $\zeta=\pi/2$, we have the following requirements
\begin{eqnarray}
w_C[\chi]+w_C[\xi]= \mbox{even number}
\end{eqnarray}
for the single-valued condition for the spin function $\Sigma_1$ as a function of the coordinate ${\bf r}$; if this condition is satisfied, 
  the product of the phase factors $e^{-{ i \over 2} \chi({\bf r})}$ and $e^{\pm i { 1 \over 2} \xi ({\bf r})}$ in Eq.~(\ref{spin-d1}) becomes single-valued.
 
 The condition $w_C[\chi]=\pm1$ requires that $w_C[\xi]$ must be odd, thus, $w_C[\xi]$ is not zero. The nonzero value of $w_C[\xi]$ means that electrons perform spin-twisting itinerant motion. This indicates that the quantized cyclotron motion without an external magnetic field is possible when the itinerant motion is accompanied by the spin-twisting. 

 \section{Stabilization of the spin-twisting itinerant motion of electrons by electron-pair formation}
 \label{section9}
 
 We explain the appearance of ${\bf A}^{\rm MB}$ by the pair formation of spin-twisting itinerant electrons in this section.
 We consider the case where the Rashba interaction $H_{so}$ which is much smaller than the pairing energy gap ( $|{\bm \lambda}| \ll |\Delta|$) is added.
 In this case, the pairing is modified from the $({\bf k}, \uparrow)$ and $(-{\bf k}, \downarrow)$ pairing to $({\bf k}, {\bf s}_0({\bf r}))$ and $(-{\bf k}, -{\bf s}_0({\bf r}))$, or $({\bf k}, -{\bf s}_0({\bf r}))$ and $(-{\bf k}, {\bf s}_0({\bf r}))$ pairing\cite{koizumi2020},
 where ${\bf k}$ is the wave vector and $\uparrow$, $\downarrow$, and ${\bf s}_0$ are the spin states. 
 The energy gap by the electron pairing is shown to be reduced by the factor $\exp \left( -{ {|{\bm \lambda|^2 } \over {|\Delta|^2}}} \right)$ \cite{koizumi2020}, which is almost one in the present case due to the condition $|{\bm \lambda}| \ll |\Delta|$.

We assume that the spin ${\bf s}_0({\bf r})$ depends on the coordinate ${\bf r}$, and describes the spin-twisting itinerant motion; ${\bf s}_0({\bf r})$ for the first pair, $({\bf k}, {\bf s}_0({\bf r}))$ and $(-{\bf k}, -{\bf s}_0({\bf r}))$, arises from the spin function $\Sigma_1$ in Eq.~(\ref{spin-d1}), and $- {\bf s}_0({\bf r})$ for the second pair, $({\bf k}, -{\bf s}_0({\bf r}))$ and $(-{\bf k}, {\bf s}_0({\bf r}))$, arises from the spin function $\Sigma_2$ given by
\begin{eqnarray}
 \Sigma_2({\bf r})=e^{-{ i \over 2} \chi({\bf r})} 
 \left(
 \begin{array}{c}
 ie^{-i { 1 \over 2} \xi ({\bf r})} \cos {{\zeta({\bf r})} \over 2}
 \\
 -i e^{i { 1 \over 2} \xi {\bf r})} \sin {{\zeta({\bf r})} \over 2}
 \end{array}
 \right)
 \end{eqnarray}
 Note that  $\Sigma_1$ and $\Sigma_2$ are orthogonal, thus, the four states $({\bf k}, {\bf s}_0({\bf r}))$, $(-{\bf k}, -{\bf s}_0({\bf r}))$, $({\bf k}, -{\bf s}_0({\bf r}))$, and $(-{\bf k}, {\bf s}_0({\bf r}))$ can be occupied, simultaneously.
 
The fictitious vector potential ${\bf A}_2^{\rm fic}({\bf r})$ from $\Sigma_2$ is calculated as
\begin{eqnarray}
{\bf A}_2^{\rm fic}({\bf r})=-i {\hbar \over e}\Sigma_2^{\dagger} \nabla \Sigma_2= -{\hbar \over {2e}} \nabla \chi ({\bf r}) -{ \hbar \over {2e}} \nabla \xi ({\bf r}) \cos \zeta ({\bf r})
\end{eqnarray}
 and the effective vector potential is given by
 \begin{eqnarray}
{\bf A}_2^{\rm eff} ={\bf A}^{\rm em}+ {\bf A}_2^{\rm fic}
\end{eqnarray}

 The single-particle energy for the pair $({\bf k}, {\bf s}_0({\bf r}))$ and $(-{\bf k}, -{\bf s}_0({\bf r}))$ is $ {\cal E}_{+}({\bf k}, {\bf r})$, and that for the pair $({\bf k}, -{\bf s}_0({\bf r}))$ and $(-{\bf k}, {\bf s}_0({\bf r}))$  is $ {\cal E}_{-}({\bf k}, {\bf r})$, where  ${\cal E}_{\pm} ({\bf k}, {\bf r})$ are given by 
 \begin{eqnarray}
{\cal E}_{\pm} ({\bf k}, {\bf r})={\cal E}({\bf k}) \pm\hbar {\bm \lambda} ({\bf r}) \times {\bf k} \cdot {\bf s}_0({\bf r})
\label{Epm}
\end{eqnarray}

Let us specify ${\bf s}_0$. Note that ${\bf k}$ is related to the wave vector ${\bf q}$ in the periodic lattice as 
\begin{eqnarray}
{\bf k} = {\bf q}+{ e \over {\hbar}} {\bf A}^{\rm eff}_1({\bf r}),  \quad {\bf k} = {\bf q}+{ e \over {\hbar}} {\bf A}^{\rm eff}_2 ({\bf r})
\label{gaugek}
\end{eqnarray}

We choose ${\bf s}_0$ to satisfy
\begin{eqnarray}
{\bm \lambda ({\bf r})} \times {\bf s}_0({\bf r})  \parallel {\bf A}_1^{\rm eff}({\bf r}) \mbox{ and }  {\bf A}_2^{\rm eff}({\bf r}) 
\label{conds0}
\end{eqnarray}
from the energy minimization condition.

For simplicity, we assume ${\bm \lambda ({\bf r})}$ is in the $z$-direction; then, the optimal ${\bf s}_0({\bf r})$ that satisfies the above condition lies in the $xy$ plane. 
Then,
$\zeta$ in ${\bf A}_1^{\rm eff}({\bf r})$ and ${\bf A}_2^{\rm eff}({\bf r})$ are $\zeta=\pi/2$, yielding
\begin{eqnarray}
{\bf A}^{\rm fic}({\bf r})=-{\hbar \over {2e}} \nabla \chi={\bf A}_1^{\rm fic}({\bf r})={\bf A}_2^{\rm fic}({\bf r})
\label{eqAAfic}
\end{eqnarray}

As a consequence, we have the common effective potential given by
\begin{eqnarray}
{\bf A}^{\rm eff}={\bf A}^{\rm em}+{\bf A}^{\rm fic}
\end{eqnarray}
for ${\bf A}_1^{\rm eff}({\bf r})$ and ${\bf A}_2^{\rm eff}({\bf r})$.

It can be shown that for the pairing of time-reversal partner states, the kinetic energy increase by the appearance of ${\bf A}^{\rm eff}$ is calculated as 
 \begin{eqnarray}
 \int d^3 r  {{e^2 n({\bf r})} \over {2m}} |{\bf A}^{\rm eff}|^2 
 \end{eqnarray}
 assuming the time and/or space inversion symmetry \cite{koizumi2020}.
 Then, the optimum ${\bf A}^{\rm fic}$ is the one that gives ${\bf A}^{\rm eff}=0$ if this choice is possible. 
 
 We may adopt ${\bf A}^{\rm em}=0$ when a magnetic field is zero. Then, the condition ${\bf A}^{\rm eff}=0$ leads to the condition ${\bf A}^{\rm fic}=0$, i.e., the absence of the spin-twisting itinerant motion.
 When ${\bf A}^{\rm em} \neq 0$, however, the optimal ${\bf A}^{\rm fic}$ will be the one for the presence of the spin-twisting itinerant motion.

Let us consider the case where the many-body wave function$\Psi$ is given as a Slater determinant
of spin-orbitals 
\begin{eqnarray}
\phi_{1,1}({\bf r})\Sigma_1({\bf r}), \  \phi_{1,2}({\bf r})\Sigma_2({\bf r}), \dots, \phi_{{N \over 2},1}({\bf r})\Sigma_1({\bf r}), \ \phi_{{N \over 2},2}({\bf r})\Sigma_2({\bf r}) 
\end{eqnarray}
where $\phi_{j,1}({\bf r})$ and $\phi_{j,2}({\bf r})$ are orbital functions; they are time-reversal partners. The total number of electrons $N$ is assumed to be even. 

Then, ${\bf A}^{\rm MB}$ is calculated as
\begin{eqnarray}
 {\bf A}^{\rm MB}&=& \  {\mathscr Im} \left\{
{ { \sum_{j=1}^{N \over 2} \left[ \phi^{\ast}_{j,1}({\bf r})\Sigma^{\dagger}_1({\bf r}) \nabla \phi_{j,1}({\bf r})\Sigma_1({\bf r})+\phi^{\ast}_{j,2}({\bf r})\Sigma^{\dagger}_2({\bf r})\nabla 
 \phi_{j,2}({\bf r})\Sigma_2({\bf r}) \right]} \over
  { \sum_{j=1}^{N \over 2} \left[ \phi^{\ast}_{j,1}({\bf r}) \phi_{j,1}({\bf r})+\phi^{\ast}_{j,2}({\bf r})
 \phi_{j,2}({\bf r}) \right] } } \right\}
 \nonumber
 \\
 &=&
 \  { e \over \hbar}
{ {{\bf A}^{\rm fic}_1\sum_{j=1}^{N \over 2}\phi^{\ast}_{j,1}({\bf r})
 \phi_{j,1}({\bf r}) + {\bf A}^{\rm fic}_2\sum_{j=1}^{N \over 2}\phi^{\ast}_{j,2}({\bf r})
 \phi_{j,2}({\bf r})} \over
  { \sum_{j=1}^{N \over 2} \left[ \phi^{\ast}_{j,1}({\bf r}) \phi_{j,1}({\bf r})+\phi^{\ast}_{j,2}({\bf r})
 \phi_{j,2}({\bf r}) \right] } }
\end{eqnarray}
where
the fact that $\sum_{j=1}^{N \over 2}[ \phi^{\ast}_{j,1}({\bf r}) \nabla \phi_{j,1}({\bf r})$+$\phi^{\ast}_{j,2}({\bf r})\nabla \phi_{j,2}({\bf r})]$ is real (due to the fact that $\phi_{j,1}({\bf r})$ and $\phi_{j,2}({\bf r})$ are time-reversal partners) is used.

The optimal ${\bf A}^{\rm fic}_1$ and ${\bf A}^{\rm fic}_2$ are given in Eq.~(\ref{eqAAfic}).
Thus,  we have
\begin{eqnarray}
 {\bf A}^{\rm MB}={ e \over \hbar} {\bf A}^{\rm fic}=-{ 1 \over 2} \nabla \chi
 \label{AMB}
\end{eqnarray}

Then, we may identify 
\begin{eqnarray}
\theta=-\chi={ 2 \over \hbar}\chi_s
\end{eqnarray}

Now the origin of $\theta$ or $\chi_s$ is identified as the Berry connection.
 
\section{${\bf A}^{\rm fic}$ from spin-vortices created by itinerant electrons in a two-dimensional model}

So far, we have considered the appearance of ${\bf A}^{\rm fic}$ in band electrons.
In this section, we consider electrons in a doped Mott insulator. 
The parent compounds of hole-doped cuprate superconductors are believed to be Mott insulators well-described by the half-filled Hubbard model with large on-site Coulomb repulsion $U$ compared with the transfer integral $t$. 
We will explain the appearance of ${\bf A}^{\rm fic}$ in a model that has been used by us as a very crude model for the bulk of the cuprate superconductors. 
In the bulk, the effect of small polaron formation is important in addition to the strong on-site Coulomb repulsion.
Due to the small polaron formation, we assume that the bulk electronic state is in the effectively-half filled situation since the mobility of the small polaron is very small at low temperatures. Then, the electronic state calculation becomes significantly easier than the ordinary Hubbard model, and electronic states with spin-twisting itinerant motion of electrons can be obtained in a numerically tractable way. 
A different model is needed in the surface region where the small polaron formation is expected to be suppressed, but we consider here only the bulk part of the problem.

\begin{figure}
\begin{center}
\includegraphics[width=8.0cm]{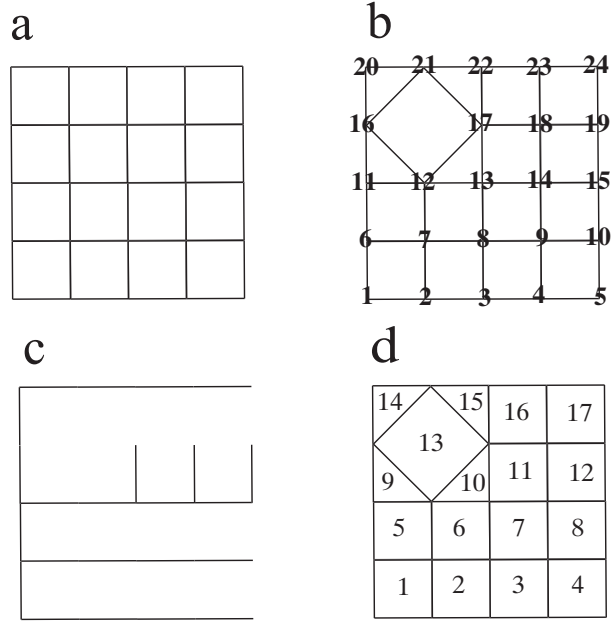}
\end{center}
\caption{Two dimensional square lattice. {\bf a}) $5\times 5$ lattice with 25 sites. {\bf b}) A lattice with 24 sites, where one site is removed from {\bf a} as it is occupied by a localized hole (small polaron). Four bonds are added around the localized hole. {\bf c}) A singlely-connected lattice made by removing some bonds from {\bf b}.  {\bf d}) 15 `` holes'' in the region where the winding numbers of $\chi$ around them have to be supplied.}
\label{Lattice}
\end{figure}

The model Hamiltonian is the following modified Hubbard model for electrons in the two-dimensional square lattice 
\begin{eqnarray}
 H
  \!&=&\!-t\sum_{\langle i,j \rangle_{1},\sigma}(c^\dagger_{i\sigma}c_{j\sigma}+c^\dagger_{j\sigma}
   c_{i\sigma})
  \!+\!U\sum_{j} c^{\dagger}_{j\uparrow} c_{j\uparrow} c^{\dagger}_{j \downarrow}c_{j\downarrow}
  \!+\!J_h \sum_{\langle i,j \rangle_{h}} \hat{\bf S}_{i}\cdot \hat{\bf S}_{j}\!+\!H^{\rm 2D}_{\rm so}
  \nonumber
  \\
  \label{Ham}
  \end{eqnarray}
   Here, $c^{\dagger}_{j \sigma}$ and  $c_{j \sigma}$ are the creation and annihilation operators of electrons at the $j$th site with the $z$-axis projection of electron spin $\sigma$, respectively, where $z$-direction is normal to the two dimensional plane.  This describes the CuO$_2$ plane of the bulk of the cuprate superconductor, assuming that small polarons are formed from the doped holes.
  Only Cu's at the lattice sites are explicitly taken into account. In Fig.~\ref{Lattice}{\bf a}, the $5 \times 5$ lattice is depicted.
  
 The first two terms in Eq.~(\ref{Ham}) come form the Hubbard Hamiltonian, where $\langle i, j \rangle_1$ indicates the nearest neighbor site pairs. The parameter $t$ is the transfer integral and it is taken as the units of energy. $U$ is the on-site Coulomb parameter; we use $U=8t$ in the present work.
  The third term describes the antiferromagnetic exchange interaction across the hole occupied sites, and $J_h$ is the coupling constant for it; we use $J_h=0.25t$ in the present work.
  
 For the sum over $i$ and $j$, hole occupied sites are excluded assuming that they are practically immobile at low temperatures due to the small polaron formation.  In Fig.~\ref{Lattice}{\bf b}, a site is removed from the lattice assuming that it is occupied by an immobile polaron.
 The sum over $\langle i, j \rangle_h$ is the sum over the pairs across the hole occupied sites, including the right angle directions. 
 The spin operator at the $j$th site $\hat{\bf S}_j$ is given by 
 \begin{eqnarray}
\hat{\bf S}_j={ 1 \over 2} \sum_{\sigma, \sigma' } c_{j \sigma}^{\dagger} {\bm \sigma}_{\sigma \sigma'} c_{j \sigma'}.
\end{eqnarray}
where ${\bm \sigma}$ is the vector of Pauli matrices.
 
  The fourth term $H^{\rm 2D}_{\rm so}$ is that for the Rashba spin-orbit interaction given by
  \begin{eqnarray}
  H^{\rm 2D}_{\rm so}
 &=&\! \lambda\sum_{h} \Big[ c^{\dagger}_{h\!+\!y \downarrow}c_{h\!-\!x \uparrow}\!-\!c^{\dagger}_{h\!+\!y \uparrow}c_{h\!-\!x \downarrow}\!+\!i(c^{\dagger}_{h\!+\!y\downarrow}c_{h\!-\!x\uparrow}\!+\!c^{\dagger}_{h\!+\!y\uparrow} c_{h\!-\!x\downarrow})
\nonumber
\\
\!&+&\! c^{\dagger}_{h\!+\!x \downarrow}c_{h\!-\!y \uparrow}\!-\!c^{\dagger}_{h\!+\!x \uparrow}c_{h\!-\!y \downarrow}\!+\!i(c^{\dagger}_{h\!+\!x\downarrow}c_{h\!-\!y\uparrow}\!+\!c^{\dagger}_{h\!+\!x\uparrow} c_{h\!-\!y\downarrow})+ {\rm h.c.}\Big]
  \end{eqnarray}
 where $h$ describes the hole occupied sites \cite{Rashba2013,Koizumi2017}. 
 We restrict that holes do not come nearby due to the Coulomb repulsion. $h+ x$ ($h-x$) are nearest neighbor sites of $h$ in the $x$ direction (in the $-x$ direction); and  $h+y$ ($h-y$) are those in the $y$ direction (in the $-y$ direction). 
The Rashba interaction is assumed to exist only around the holes (along the second nearest neighbor hopping paths (see Fig.~\ref{Lattice}{\bf b})) with the internal electric field in the direction perpendicular to the CuO$_2$ plane. $ \lambda$ is the parameter for the Rashba interaction; we use   $\lambda=-0.02t$ in this work. This Rashba interaction is the minimal one, but may not be sufficient.

Since the hole occupied sites are excluded from the accessible sites,
the electron system is in the situation where the number of electrons and that of the accessible sites are equal.
We call this situation, the {\em effectively-half filled situation} (EHFS). 
The ordinary current generation by single-particle excitations is not effective due to the large $U$ value.
However, a collective motion with the effective transfer integral $t^3/U^2$ is possible when spin-vortices are created \cite{Koizumi2017}. 
Furthermore, this current is a topologically-protected one generated by ${\bf A}^{\rm fic}$. We call it, the {\em spin-vortex-induced loop current} (SVILC) \cite{HKoizumi2013,HKoizumi2014}. 
 
The many-body Hamiltonian in Eq.~(\ref{Ham}) looks simple; however, it is already too difficult to solve as it is.  Therefore, we use the following mean field version,
\begin{eqnarray}
&&H^{HF}_{EHFS}=-t\sum_{\langle i,j \rangle_{1},\sigma}(c^\dagger_{i\sigma}c_{j\sigma}+c^\dagger_{j\sigma}
   c_{i\sigma}) +J_h\sum_{\langle i,j \rangle_{h}} \left( {\bf S}_{i}\cdot\hat{{\bf S}}_{j}+{\bf S}_{j}\cdot\hat{{\bf S}}_{i} -{\bf S}_{j}\cdot{{\bf S}}_{i}\right)
   +H^{\rm 2D}_{\rm so}
   \nonumber
   \\
   &&
  +U\sum_{j}\Big[ (\frac{1}{2}-S^z_j)c^\dagger_{j\uparrow}c_{j\uparrow}+
   (\frac{1}{2}+S^z_j)c^\dagger_{j\downarrow}c_{j\downarrow}
   -(S^x_j-iS^y_j)c^\dagger_{j\uparrow}c_{j\downarrow}
   -(S^x_j+iS^y_j)c^\dagger_{j\downarrow}c_{j\uparrow}-{2 \over 3}{\bf S}^2 \Big]
   \nonumber
   \\
   \label{hhf}
\end{eqnarray}
and  ${S}^{x}_j, S^{y}_j$ and $S^{z}_j$ are expectation values of the components of $\hat{\bf S}_j$ calculated as
\begin{eqnarray}
 {S}^{x}_j&=&\frac{1}{2}\langle c^{\dagger}_{j\uparrow} c_{j\downarrow}+c^{\dagger}_{j\downarrow} c_{j\uparrow}\rangle
 =S_j\cos\xi_j\sin\zeta_j
 \nonumber
 \\
 {S}^{y}_j&=&\frac{i}{2}\langle-c^{\dagger}_{j\uparrow}\ c_{j\downarrow}+c^{\dagger}_{j\downarrow} c_{j\uparrow}\rangle
 =S_j\sin\xi_j\sin\zeta_j
 \nonumber
 \\
 {S}^{z}_j&=&\frac{1}{2}\langle c^{\dagger}_{j\uparrow} c_{j\uparrow}-c^{\dagger}_{j\downarrow} c_{j\downarrow}\rangle
 =S_j\cos\zeta_j
 \label{s3eq3}
\end{eqnarray}
with $\langle \hat{O} \rangle$ denoting the expectation value of the operator $\hat{O}$. 

Through the self-consistent calculation using $H^{HF}_{EHFS}$, we obtain the following Hartree-Fock orbitals,
 \begin{eqnarray}
 |\tilde{\gamma_k} \rangle=\sum_{j} [ \tilde{D}_{j \uparrow}^{\gamma_k} c^{\dagger}_{j \uparrow}+\tilde{D}_{j \downarrow}^{\gamma_k} c^{\dagger}_{j \downarrow}]  | {\rm vac} \rangle.
 \end{eqnarray}
where $\tilde{D}^{\gamma}_{j \sigma}$'s are numerically obtained parameters.
 
Actually,  $H^{\rm 2D}_{\rm so}$ does not contribute at all for the evaluation of $\tilde{D}^{\gamma}_{j \sigma}$'s which is done solely by energy minimization since the energy minimization procedure gives a currentless state (known as ``Bloch's Theorem" \cite{Bohm1949}). 

The Hartree-Fock orbitals $\{ |\tilde{\gamma}_k \rangle \}$ satisfy the orthonormal condition
 \begin{eqnarray}
\langle\tilde{\gamma}_j  |\tilde{\gamma}_k \rangle =\delta_{j k},
 \end{eqnarray}
 and a tentative total wave function $|\tilde{\Psi} \rangle$ is constructed as the Slater determinant of the occupied $|\tilde{\gamma}_k \rangle $'s.
 
\begin{figure}
\begin{center}
\includegraphics[width=12.0cm]{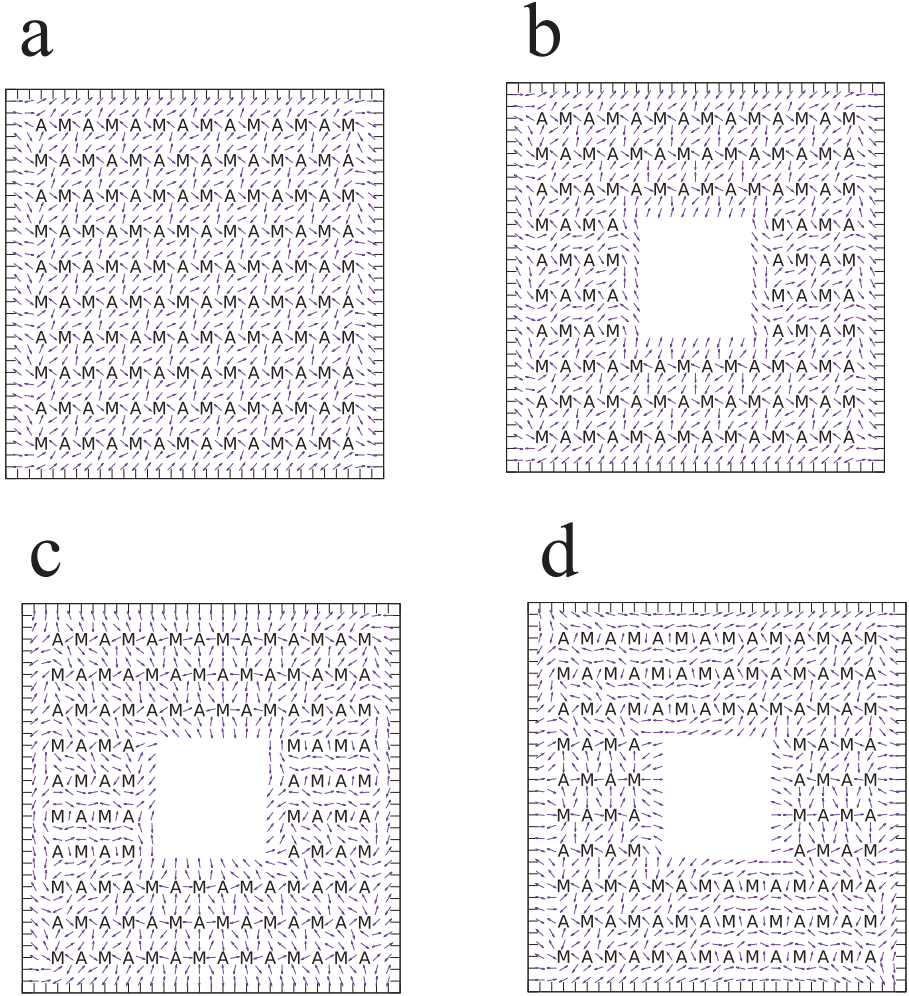}
\end{center}
\caption{Spin textures for systems with spin-vortices created by spin-twisting itinerant motion of electrons. ``M'' and ``A'' indicates centers of spin-vortices with winding numbers $+1$ and $-1$, respectively. 
{\bf a}) a case without a hole region.  {\bf b}) a case with a hole region, and the winding number for spins around the hole is zero. 
 {\bf c}) a case with a hole region, and the winding number for spins around the hole is $+1$. 
  {\bf d}) a case with a hole region, and the winding number for spins around the hole is $-1$. }
\label{xi}
\end{figure}

 In Fig.~\ref{xi}, examples of spin textures for systems with spin-vortices are depicted. Two lattice systems, one is a $31 \times 32$ lattice ({\bf a}), and the other is $31 \times 32$ lattice with a $9 \times 10$ hole region ({\bf b}-{\bf d}) are shown. 
 
Next, we calculate $\xi_j$ and $\zeta_j$. $\zeta_j$ is taken to be $\zeta_j=\pi/2$ by anticipating the energy gain by the Rashba interaction.
From the expectation value of the spin components $S^x_j$ and $S^y_j$, $\xi_j$'s are calculated using Eq.~(\ref{s3eq3}). The value of $\xi_j$ has an ambiguity of an integral multiple of $2\pi$. 
We choose a particular branch for $\xi_j$. 
Since the antiferromagnetic background from the Hubbard Hamiltonian given by $\xi^0_{j}=\pi (j_x+j_y)$ exists ($(j_x, j_y)$ is the $xy$ coordinates of the $j$th site taking the lattice constant $a=1$), we separate the antiferromagnetic contribution from $\xi$ and introduce angular variable $\eta$, 
 \begin{eqnarray}
 \eta_j=\xi_j-\pi (j_x+j_y)
 \label{eta}
 \end{eqnarray}
  where $\eta_j$ is $\eta$ at the $j$th site.
 We take the branch of $\eta_j$ that satisfies the difference of value from the nearest neighbor site $k$ is in the range,
\begin{eqnarray}
-\pi \le \eta_{j} -\eta_k < \pi
\end{eqnarray}

From ($\eta_j-\eta_k$)'s, we construct  ($\xi_j-\xi_k$)'s. After ($\xi_j-\xi_k$)'s are obtained, we rebuild $\xi$ from them. 
The process is as follows: first, we pick a value for the initial $\xi_1$  (say $\xi_1=0$). 
After fixing the value of $\xi_1$, we calculate $\xi_2$ by $\xi_{2} = \xi_1 + (\xi_{2} -\xi_1)$, where the site $2$ is connected to the site $1$ by a nearest neighbor bond.
 The step where value $\xi_{j}$ is derived from the already evaluated value of $\xi_k$ is given by
\begin{eqnarray}
\xi_{j} = \xi_k + (\xi_{j} -\xi_k)
\end{eqnarray}
where the sites ${j}$ and $k$ are connected by a bond in the path for the rebuilding of $\xi$. 
This process is continued until values at all accessible sites are evaluated once and only once. 

By this rebuilding process, a single path is constructed from the site $1$ to other sites $k\neq 1$, which is achieved by making the region singly-connected by removing some bonds as indicated in Fig.~\ref{Lattice}{\bf c}. This procedure corresponds to inserting dividing surfaces in Fig.~\ref{London-holes}.

We denote the path from the site $1$ to other sites $k\neq 1$ by $C_{1 \rightarrow k}$.
Then, the value $\xi_k$ is given as
\begin{eqnarray}
\xi_k \approx \xi_1+ \int_{C_{1 \rightarrow k}} \nabla \xi \cdot d{\bf r}
\label{rebuilxi}
\end{eqnarray}

The presence of spin-vortices are described by non-zero winding numbers, $w_{C_{\ell}}[\xi]$, where the winding number of $\xi$ for loop $C_{\ell}$ that encircles a hole is defined by
\begin{eqnarray}
w_{C_{\ell}}[\xi]={ 1 \over {2\pi}} \sum_{i=1}^{N_{\ell}} ( \xi_{C_{\ell}(i+1)} -\xi_{C_{\ell}(i)}) \approx { 1 \over {2\pi}} \oint_{C_{\ell}} \nabla \xi \cdot d{\bf r}
\label{wnumber}
\end{eqnarray}
where $N_{\ell}$ is the total number of sites on the loop $C_{\ell}$, and ${C_{\ell}(i)}$ is the $i$th site on it with the periodic condition ${C_{\ell}(N_{\ell}+1)=C_{\ell}(1)}$. Examples of such loops around holes are seen in Fig.~\ref{Lattice}{\bf d}, where $15$ loops are depicted.

The angle $\xi$ may have jump-of-values (integer multiple of $2\pi$) between bonds that are not used in the process of rebuilding its value. This jump-of-value causes
the multi-valuedness in $|\tilde{\gamma} \rangle$, since $|\tilde{\gamma} \rangle$ is actually expressed as
 \begin{eqnarray}
 |\tilde{\gamma} \rangle=\sum_{j} \left[ e^{-i{{\xi_j} \over 2}}{D}_{j \uparrow}^{\gamma} {c}^{\dagger}_{j \uparrow}+ e^{i{{\xi_j} \over 2}}{D}_{j \downarrow}^{\gamma} {c}^{\dagger}_{j \downarrow} \right]  | {\rm vac} \rangle
  \label{tabhenkan2}
 \end{eqnarray}
 It contains factors $e^{\pm i{{\xi_j} \over 2}}$ that become multi-valued in the presence of the spin-vortices.
 
Due to the multi-valuedness of $ |\tilde{\gamma} \rangle$, $|\tilde{\Psi} \rangle$ becomes multi-valued.
On the other hand, the exact total wave function must be single-valued as a function of electron coordinates. We remedy this discrepancy by adding a phase factor that 
compensates the multi-valuedness of the basis $\{ |\tilde{\gamma} \rangle \}$. Namely, 
we construct the single-valued basis $\{ |{\gamma} \rangle \}$ given by
  \begin{eqnarray}
 |\gamma \rangle& =&\sum_{j} e^{\!-\!i { {\chi_j } \over 2}} [e^{-i{{\xi_j} \over 2}}D_{j \uparrow}^{\gamma} {c}^{\dagger}_{j \uparrow}\!+\!e^{i{{\xi_j} \over 2}}D_{j \downarrow}^{\gamma} {c}^{\dagger}_{j \downarrow}] | {\rm vac} \rangle
\label{Cwave}
 \end{eqnarray}
 and obtain the single-valued total wave function $|{\Psi} \rangle$ as the Slater determinant of the occupied $|{\gamma}_k \rangle $'s.
 
 The phase factor $e^{\!-\!i { {\chi_j } \over 2}}$ plays the same role as that added in Eq.~(\ref{spin-d1}). It is specified by the condition that $|{\gamma}_k \rangle $'s are single-valued as functions of the coordinate.

 Before obtaining $\chi$, we need to evaluate ${D}^{\gamma}_{j \sigma}$'s that are compatible with the rebuilt $\xi$ obtained using $C_{1 \rightarrow k}$'s, where
 the compatible ${D}^{\gamma}_{j \sigma}$'s means they are obtained for $\xi$ whose jump-of-value locations are known.
  It is important that $\chi$ has the same jump-of-value locations.
  
  In order to obtain ${D}^{\gamma}_{j \sigma}$'s whose jump-of-value positions are known, we diagonalize the following Hamiltonian {\em one time} using ${\bf S}_j$ and $n_j$ already obtained by the self-consistent calculation;
\begin{eqnarray}
&&\tilde{H}^{HF}_{EHFS}=-t\sum_{\langle i,j \rangle_{1}}\left(e^{i {1 \over2}(\xi_i -\xi_j)} \tilde{c}^\dagger_{i\uparrow} \tilde{c}_{j\uparrow}+
e^{-i {1 \over2}(\xi_i -\xi_j)} \tilde{c}^\dagger_{i\downarrow} \tilde{c}_{j\downarrow} + {\rm H.c.} \right)
\nonumber
\\
&&  +J_h\sum_{\langle i,j \rangle_{h}} \left( {\bf S}_{i}\cdot\hat{{\bf S}}_{j}+{\bf S}_{j}\cdot\hat{{\bf S}}\right)
   \nonumber
   \\
  &&+U\sum_{j}\Big[ (\frac{1}{2}-S^z_j)\tilde{c}^\dagger_{j\uparrow} \tilde{c}_{j\uparrow}\!+\!
   (\frac{1}{2}\!+\!S^z_j) \tilde{c}^\dagger_{j\downarrow} \tilde{c}_{j\downarrow}
   \!-\!(S^x_j \!-\! iS^y_j)\tilde{c}^\dagger_{j\uparrow}\tilde{c}_{j\downarrow}
   \!-\!(S^x_j\!+\!iS^y_j)\tilde{c}^\dagger_{j\downarrow} \tilde{c}_{j\uparrow}\Big]
   \nonumber
   \\
   \label{tildhhf}
\end{eqnarray}
where
\begin{eqnarray}
\tilde{c}^{\dagger}_{j\uparrow}= {c}^{\dagger}_{j\uparrow} e^{-i {1 \over2} \xi_i}, \  \tilde{c}_{j\uparrow}= {c}_{j\uparrow} e^{i {1 \over2} \xi_i}, \ 
 \tilde{c}^{\dagger}_{j\downarrow}= {c}^{\dagger}_{j\downarrow}e^{i {1 \over2} \xi_i}, \  \tilde{c}_{j\downarrow}= {c}_{j\downarrow}e^{-i {1 \over2} \xi_i}
\end{eqnarray}
Now, using the ${D}^{\gamma}_{j \sigma}$'s obtained above, the single-particle wave functions $|\tilde{\gamma} \rangle$ is given by
\begin{eqnarray}
 |\tilde{\gamma} \rangle=\sum_{j} \left[{D}_{j \uparrow}^{\gamma} \tilde{c}^{\dagger}_{j \uparrow}+ {D}_{j \downarrow}^{\gamma} \tilde{c}^{\dagger}_{j \downarrow} \right]  | {\rm vac} \rangle
 \end{eqnarray}

When spin-vortices are present, phase factors $e^{\pm i\frac{\xi_j}{2}}$ in Eq.~(\ref{Cwave}), become multi-valued with respect to the coordinate since
$\xi_j$ has ambiguity of adding an integral multiple of $2\pi$.
 To restore the single-valuedness, the phase $\chi$ satisfies the following condition,
 \begin{eqnarray}
w_{C_{\ell}}[\xi]+ w_{C_{\ell}}[\chi] = \mbox{even number}  \mbox{  for any loop $C_{\ell}$}
\label{windcond}
\end{eqnarray}

The angular variable $\chi$ for the ground state is obtained by minimizing the total energy by imposing the above constraint.
We obtain $(\chi_k-\chi_j)$'s by minimizing the following functional 
\begin{eqnarray}
F[\nabla \chi]=E[\nabla \chi]+\sum_{\ell=1}^{N_{\rm loop}} { {\lambda_{\ell}}}\left(  \oint_{C_\ell} \nabla \chi \cdot d {\bf r}-2 \pi w_{C_{\ell}}[\chi] \right), 
\label{functional}
\end{eqnarray}
where 
\begin{eqnarray}
E[\nabla \chi]=\langle {\Psi} | H^{HF}_{EHFS} |{\Psi} \rangle
\label{energyf}
\end{eqnarray}
 $\lambda_{\ell}$'s are Lagrange multipliers, and $\{ C_1, \cdots, C_{N_{\rm loop}} \}$ are boundaries of plaques of the lattice, where $N_{\rm loop}$ is equal to the
number of plaques of the lattice. In Fig.~\ref{Lattice}{\bf d},  a $N_{\rm loop}=15$ case is depicted.
  
We take the branch of $\chi_j$ that satisfies the difference of value from the nearest neighbor site $k$ is in the range,
\begin{eqnarray}
-\pi \le \chi_{j} -\chi_k < \pi
\end{eqnarray}
 
We rebuild $\chi$ from  $(\chi_k-\chi_j)$'s in a similar manner as $\xi$ is rebuilt. Thus, $\chi_k$ is given by
\begin{eqnarray}
\chi_k \approx \chi_1+ \int_{C_{1 \rightarrow k}} \nabla \chi \cdot d{\bf r}
\label{rebuilchi}
\end{eqnarray}

$(\chi_k-\chi_j)$'s are obtained as solutions of the following system of equations;
\begin{eqnarray}
{{\delta E[\nabla \chi]} \over {\delta \nabla \chi}}&+&\sum_{\ell=1}^{N_{\rm loop}} { {\lambda_{\ell}}} {{\delta } \over {\delta \nabla \chi}} \oint_{C_\ell} \nabla \chi \cdot d {\bf r}=0
\label{Feq1}
\\
 \oint_{C_\ell} \nabla \chi \cdot d {\bf r}&=&2 \pi w_{C_{\ell}}[\chi]
 \label{Feq2}
\end{eqnarray}
 A set of parameters $\{ w_{C_{\ell}}[\chi] \}$ must be supplied as part of boundary conditions. The number of them is $N_{\rm loop}$, which is equal to the number of 
 $\{ \lambda_{\ell} \}$ to be evaluated.

In the discrete lattice, Eqs.~(\ref{Feq1}) and (\ref{Feq2}) are given by
\begin{eqnarray}
{{\partial  E( \{ \tau_{k \leftarrow j} \}) } \over {\partial \tau_{k \leftarrow j} }}&+&\sum_{\ell=1}^{N_{\rm loop}}  { {\lambda_{\ell}}} 
{{\partial } \over {\partial \tau_{k \leftarrow j}}} \sum_{k \leftarrow j} L_{k \leftarrow j}^{\ell}\tau_{k \leftarrow j} =0
\label{Feq1b}
\\
 \sum_{k \leftarrow j} L_{k \leftarrow j}^{\ell}\tau_{k \leftarrow j} &=& 2 \pi w_{C_{\ell}}[\chi]
 \label{Feq2b}
\end{eqnarray}
where the sum is taken over the bonds ${k \leftarrow j}$, $\tau_{ k \leftarrow j}$ is the difference of $\chi$ for the bond $\{k \leftarrow j \}$
\begin{eqnarray}
\tau_{ k \leftarrow j}=\chi_k -\chi_j 
\end{eqnarray}
and $L_{k \leftarrow j}^{\ell}$ is defined as
\begin{eqnarray}
L_{k \leftarrow j}^{\ell} =
\left\{
\begin{array}{cl}
-1 & \mbox{ if  $ k \leftarrow j$ exists in $C_{\ell}$ in the clockwise direction}
\\
1 & \mbox{ if  $ k \leftarrow j$ exists in $C_{\ell}$ in the counterclockwise direction}
\\
0 & \mbox{ if  $ k \leftarrow j$ does not exist in $C_{\ell}$}
\end{array}
\right.
\end{eqnarray}
The number of equations in Eqs.~(\ref{Feq1b}) and (\ref{Feq2b}) is
(the number of bonds)+(the number of plaques) which is equal to the number of unknowns $\{ \tau_{ j \leftarrow i} \}$ and $\{ \lambda^{\ell} \}$.

 \begin{figure}
\begin{center}
\includegraphics[width=11.0cm]{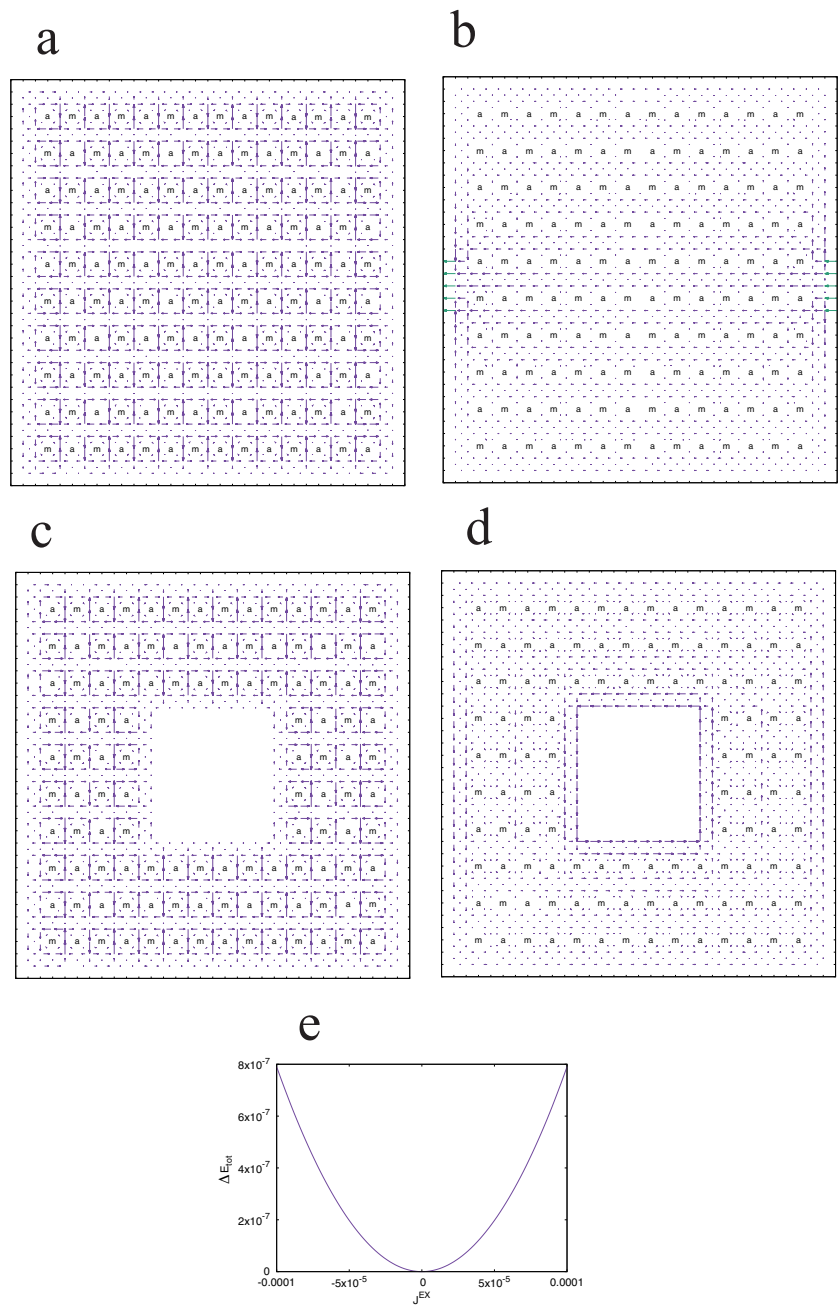}
\end{center}
\caption{Current distribution for systems with SVILCs and the total energy dependence of $J^{\rm EX}$.  ``m'' and ``a'' indicates centers of loop currents with winding numbers $+1$ and $-1$, respectively. 
{\bf a}) current distribution for the system with spin-texture in Fig.~\ref{xi}{\bf a}. The arrows are normalized so that the largest one is equal to the lattice spacing.  {\bf b}) difference of the current distribution for the system with spin-texture in Fig.~\ref{xi}{\bf a} by the external current feeding indicated by green arrows. $J^{\rm EX}$ is fed from each site.
{\bf c}) current distribution for the system with spin-texture in Fig.~\ref{xi}{\bf b}. {\bf d}) difference of the current distribution for the system with spin-texture in Fig.~\ref{xi}{\bf b} by changing the spin-winding number for the hole from $0$ to $+1$. {\bf e}) $J^{\rm EX}$ dependence of  total energy for the system with spin-texture in Fig.~\ref{xi}{\bf c} and the loop current winding number $+1$ around the hole region. The current feeding is the same as in {\bf b}.} 
\label{current}
\end{figure}

Actually, we can obtain $\tau_{ j \leftarrow i}$'s without obtaining ${ {\lambda_{\ell}}}$'s. This method is convenient when the feeding current is introduced and the Rashba spin-orbit interaction is included. 
We shall explain this method, below.

First, we note that the current density is given by
\begin{eqnarray}
{\bf J}={{2e} \over \hbar}  {{\delta E} \over {\delta \nabla \chi}}
\end{eqnarray}
For the lattice system, it is expressed as
\begin{eqnarray}
J_{ j \leftarrow i}= {{2e} \over \hbar}  {{\partial E} \over {\partial \tau_{ j \leftarrow i}}}
\end{eqnarray}
where $J_{ j \leftarrow i}$ is the current through the bond between sites $i$ and $j$ in the direction  $j \leftarrow i$.

Then, the conservation of charge at site $j$ is given by
\begin{eqnarray}
0=J^{\rm EX} _{ j} + \sum_i {{2e} \over \hbar}  {{\partial E} \over {\partial \tau_{ j \leftarrow i}}}
\label{Feq3}
\end{eqnarray}
where $J^{\rm EX} _{ j}$ is the external current that enters through site $j$. We use this in place of Eq.~(\ref{Feq1b}).

In order to impose conditions in Eq.~(\ref{Feq2b}), $\tau_{ j \leftarrow i}$ is split into a multi-valued part $\tau_{ j \leftarrow i}^0$ and 
single-valued part $f_{ j \leftarrow i}$ as
\begin{eqnarray}
\tau_{ j \leftarrow i}=\tau_{ j \leftarrow i}^0 + f_{ j \leftarrow i}
\end{eqnarray}
where $\tau_{ j \leftarrow i}^0$ satisfies the constraint in Eq.~(\ref{Feq2b})
\begin{eqnarray}
w_{C_{\ell}}[\chi]={ 1 \over {2\pi}} \sum_{i=1}^{N_{\ell}}  \tau^0_{{C_{\ell}(i+1)} \leftarrow {C_{\ell}(i)}}
\end{eqnarray}
and $f_{ j \leftarrow i}$ satisfies
\begin{eqnarray}
0={ 1 \over {2\pi}} \sum_{i=1}^{N_{\ell}}  f_{{C_{\ell}(i+1)} \leftarrow {C_{\ell}(i)}}
\end{eqnarray}

We employ an iterative improvement of the approximate solutions by using the linearized version of Eq.~(\ref{Feq3}) given by
\begin{eqnarray}
0 \approx J^{\rm EX} _{ j} + {{2e} \over \hbar} \sum_i  {{\partial E (\{ \tau_{ j \leftarrow i}^0 \}) } \over {\partial \tau_{ j \leftarrow i}}}+{{2e} \over \hbar} 
\sum_i  {{\partial^2 E (\{ \tau_{ j \leftarrow i}^0 \}) } \over {\partial (\tau_{ j \leftarrow i})^2}}f_{ j \leftarrow i}
\label{Feq4}
\end{eqnarray}
These equations are solved for $f_{ j \leftarrow i}$'s for given $\tau_{ j \leftarrow i}^0$'s. Then, $\tau_{ j \leftarrow i}^0$'s are updated at each iteration as
\begin{eqnarray}
\tau_{ j \leftarrow i}^{ 0 \ 
New}=\tau_{ j \leftarrow i}^{ 0 \ Old}+f_{ j \leftarrow i}
\end{eqnarray}
where $\tau_{ j \leftarrow i}^{ 0 \ Old}$ is $\tau_{ j \leftarrow i}^{ 0}$ value that is used to obtain the current value of $f_{ j \leftarrow i}$; $\tau_{ j \leftarrow i}^{ 0 \ New}$ will be used to obtain the next $f_{ j \leftarrow i}$ value. 

The numerical convergence is checked by the condition
\begin{eqnarray}
\left| J^{\rm EX} _{ j} + {{2e} \over \hbar} \sum_i  {{\partial E (\{ \tau_{ j \leftarrow i}^0 \}) } \over {\partial \tau_{ j \leftarrow i}}} \right| < \epsilon
\end{eqnarray}
where $\epsilon$ is a small number.

For the initial $\tau_{ j \leftarrow i}^{ 0 }$, we adopt the following,
\begin{eqnarray}
\tau_{ j \leftarrow i}^{ 0 \ init}=\sum_h  w_h \tan^{-1} {{ j_y- h_y} \over {j_x -h_x}}-\sum_h w_h \tan^{-1} {{ i_y- h_y} \over {i_x -h_x}}
\end{eqnarray}
where $(j_x,j_y)$ and $(i_x,i_y)$ are  coordinates of the sites $j$ and $i$, respectively, $h=(h_x,h_y)$ is the coordinate of the hole occupied site, and $w_h$ is the winding number of $\chi$ around the hole at $h$.

For the lattice system, the number of $\tau_{ j \leftarrow i}$ to be evaluated is equal to the number of the bonds.
The number of equations in Eq.~(\ref{Feq2b}) is equal to the number of the plaques.
This corresponds to supplying  $\Phi_k$'s in Fig.~\ref{London-holes}.
The number of equations from Eq.~(\ref{Feq3}) for the conservation of charge is equal to the number of sites$-1$, due to the fact that the total charge is fixed in the calculation.

The equality of the number of unknowns and the number of equations gives
\begin{eqnarray}
[\mbox{\# bonds}]=[\mbox{\# plaques}]+[\mbox{\# sites}-1]
\label{Euler1}
\end{eqnarray}

 It is interesting to note that this agrees with the Euler's theorem for the two-dimensional lattice 
\begin{eqnarray}
[\mbox{\# edges}]=[\mbox{\# faces}]+[\mbox{\# vertices}-1]
\label{Euler2}
\end{eqnarray}

When the current is non-zero, the total energy depends on $\xi_1$ in Eq.~(\ref{rebuilxi}) since the Rashba spin-orbit interaction depends on the relative directions of spin and current. We optimized $\xi_1$ to minimize the total energy.
 In Figs.~\ref{current}{\bf a}-{\bf d}, current distributions for systems with SVILCs are depicted. In Fig.~\ref{current}{\bf e}, $J^{\rm EX}$ dependence of the total energy is shown. 
 The total energy is minimum at zero feeding current.
 
 \begin{figure}
\begin{center}
\includegraphics[width=11.0cm]{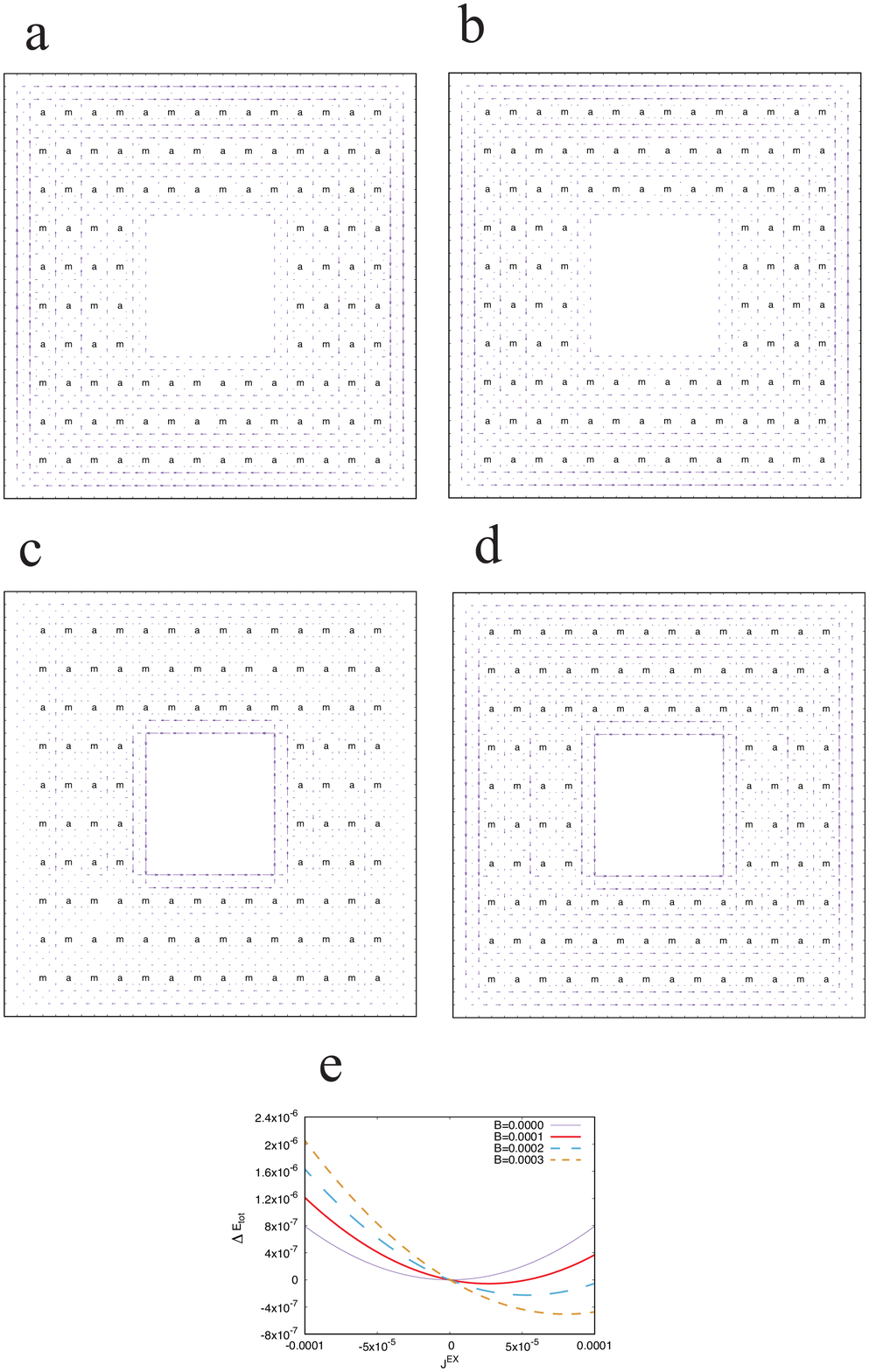}
\end{center}
\caption{Current distribution for systems with SVILCs and the total energy dependence of $J^{\rm EX}$. {\bf a}) difference of the current distribution for the system with spin-texture in Fig.~\ref{xi}{\bf b} by applying the magnetic field $B=0.01$. 
  {\bf b}) the same as {\bf a} but with $B=-0.01$.
  {\bf c}) difference of the current distribution for the system with spin-texture in Fig.~\ref{xi}~{\bf c} with the loop current winding number around the hole region $+1$ by applying the magnetic field $B=0.01$. 
  {\bf d}) the same as {\bf c} but with $B=-0.01$.
 {\bf e}) $J^{\rm EX}$ dependence of total energy for the system with spin-texture in Fig.~\ref{xi}{\bf c} and the loop current winding number $+1$ around the hole region, with varying $B$.} 
\label{currentAem}
\end{figure}
  
 Let us consider the situation where a magnetic field ${\bf B}^{\rm em}=\nabla \times {\bf A}^{\rm em}$ is applied.
Then, the energy functional  in Eq.~(\ref{energyf}) is modified as
\begin{eqnarray}
E[\nabla \chi] \rightarrow E \left[\nabla \chi -{{2e} \over \hbar}{\bf A}^{\rm em} \right]
\end{eqnarray}

This leads to replace $\tau_{ j \leftarrow i}$ in $E$ by
\begin{eqnarray}
u_{ j \leftarrow i}=\tau_{ j \leftarrow i}-{{2e} \over \hbar} \int^j_i{\bf A}^{\rm em}\cdot d{\bf r}
\end{eqnarray}
where integration is performed along the bond $ j \leftarrow i$.

The calculation can be done similarly to the case for no magnetic field, starting from the initial value
\begin{eqnarray}
u_{ j \leftarrow i}^{0 \ init}=\tau_{ j \leftarrow i}^{ 0  \ init}-{{2e} \over \hbar} \int_{j \leftarrow i}{\bf A}^{\rm em}\cdot d{\bf r}
\end{eqnarray}

Note that during the evaluation process of $\nabla \chi$, the ambiguity in the gauge of ${\bf A}^{\rm em}$ is compensated, thus, the effective vector potential 
\begin{eqnarray}
{\bf A}^{\rm eff}= {\bf A}^{\rm em}-{\hbar  \over {2e}} \nabla \chi
\end{eqnarray}
is gauge invariant with respect to the choice of the gauge in ${\bf A}^{\rm em}$. 

Now we apply a uniform magnetic field perpendicular to the lattice.
In the actual numerical calculations we have adopted
\begin{eqnarray}
{\bf A}^{\rm em}=
\left(
\begin{array}{c}
-By \\
0 \\
0
\end{array}
\right)
\end{eqnarray}
The calculated current distribution is identical even other gauge is employed.

In Figs.~\ref{currentAem}{\bf a}-{\bf d}, current distributions with applying the magnetic field are depicted.
The induced current is a diamagnetic current as is clearly shown in Fig.~\ref{currentAem}{\bf d}.
In Fig.~\ref{currentAem}{\bf e}, $J^{\rm EX}$ dependence of the total energy with applying the magnetic field are depicted.
It is important to note that the energy minimum occurs at nonzero $J^{\rm EX}$. This means that spontaneous current feeding  with the minimal  
$J^{\rm EX}$ value occurs. 

 \begin{figure}
\begin{center}
\includegraphics[width=7.0cm]{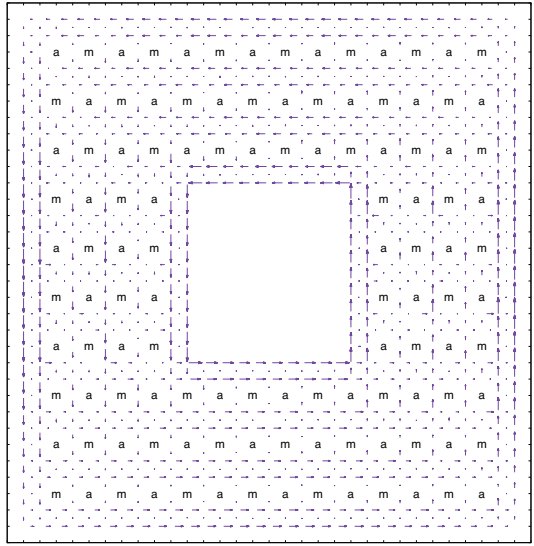}
\end{center}
\caption{Difference of current distribution for systems calculated using Eq.~(\ref{eqLondonAp}), corresponding to Fig.~\ref{currentAem}{\bf d}} 
\label{currentLondon}
\end{figure}

Numerical calculations indicate the following relation hold
\begin{eqnarray}
J_{j \leftarrow i}\approx -{{4 e^2} \over \hbar^2} { {\partial^2 E[\{0\}]} \over {\partial (u_{j \leftarrow i})^2} } \int_i^j {\bf A}^{\rm eff} \cdot d {\bf r}
\label{eqLondonAp}
\end{eqnarray}
where $\{0\}$ means all $u_{j \leftarrow i}$'s are zero \cite{Manabe2019}. This corresponds to the London formula in Eq.~(\ref{eqLondon0}).
An example calculation using the above approximation is shown in Fig.~\ref{currentLondon}.
The approximate result is almost identical to the exact one, except the current just around the small polarons that originates directly from the Rashba interaction.

 \begin{figure}
\begin{center}
\includegraphics[width=11.0cm]{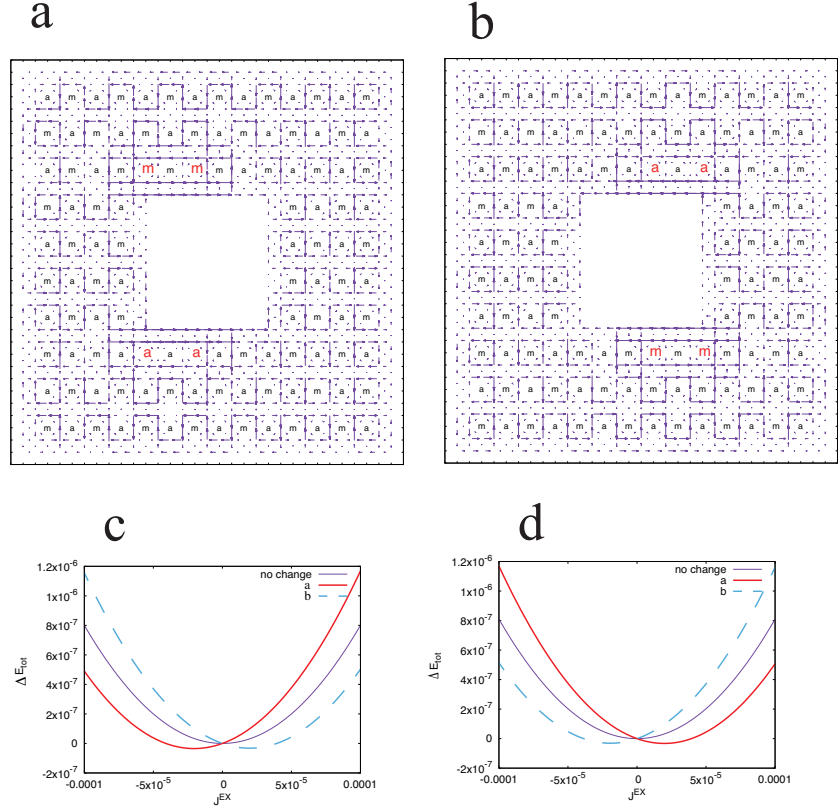}
\end{center}
\caption{Current distribution for systems with SVILCs and the total energy dependence of $J^{\rm EX}$. Larger ``m'' and ``a'' indicate the altered loop current winding numbers from the original one depicted in Fig.~\ref{xi}{\bf c}. {\bf a}) current distribution for the system with spin-texture with four altered loop current winding numbers. 
{\bf b}) the same as {\bf a} but with different alteration.
{\bf c}) $J^{\rm EX}$ dependence of total energy for the system with spin-texture in Fig.~\ref{xi}{\bf b} and the loop current winding number $0$ around the hole region. {\bf d}) total energy dependence of $J^{\rm EX}$ for the system with spin-texture in Fig.~\ref{xi}{\bf c} and the loop current winding number $+1$ around the hole region.} 
\label{currentSP}
\end{figure}

In Figs.~\ref{currentSP}{\bf a}-{\bf b}, current distributions with some of the SVILCs winding numbers flipped are shown. 
In Fig.~\ref{currentSP}{\bf c}-{\bf d}, $J^{\rm EX}$ dependence of the total energy in the SVILCs winding number flipped states are depicted. In this case, the energy minimum occurs at nonzero $J^{\rm EX}$ even without magnetic field. 
These states can be identified as the superconducting states with spontaneus external current feeding. These state explain the zero voltage current flow through superconductors.
If we turn-off the Rashba interaction, the minimum becomes at zero $J^{\rm EX}$. This indicates that the appearance of the minimum at $J^{\rm EX} \ne 0$ is due to the presence of the Rashba interaction.

\section{Concluding remarks}

We have shown that supercurrent arises when spin-twisting itinerant motion is realized and the nontrivial Berry connection appears. The supercurrent is given as a collection of topologically-protected loop currents.
For the appearance of the supercurrent, the Rashba interaction is necessary. 
The gauge invariant supercurrent arises without the breakdown of the global $U(1)$ gauge invariance.

\begin{figure}
\begin{center}
\includegraphics[scale=0.4]{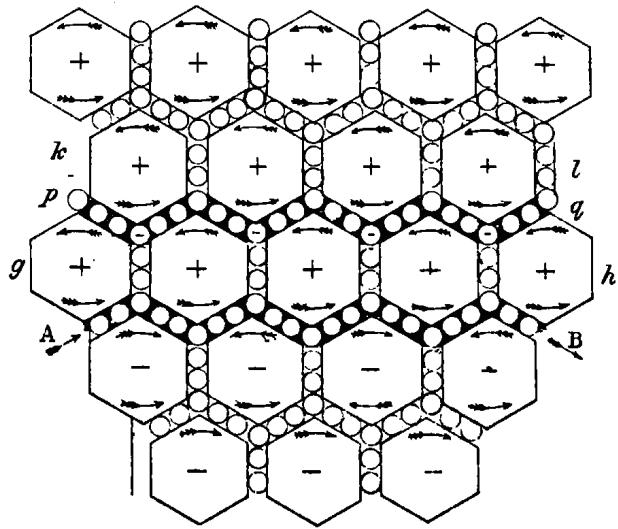}
\end{center}
\caption{The induced current generation by a magnetic field envisaged by Maxwell. This figure is taken from Ref.~\onlinecite{Maxwell2} (public domain). Small circles represent idle wheels. ``+'' and ``-'' indicate the lines of force (associated with the magnetic field) by the molecular vortices. The current enters from $A$ and exits to $B$.}  
\label{Maxwell}
\end{figure}

The topologically-protected loop current presented in this work resembles somewhat to the ``idle wheel'' envisaged by Maxwell in order to produce electric current by an application of a magnetic field \cite{Maxwell2}.
In Fig.~\ref{Maxwell}, the induced current generation by a magnetic field envisaged by Maxwell is depicted.
The following association should not take, literally; it is a very figurative one.

Each circle in Fig.~\ref{Maxwell} indicates an ``idle wheel'', which seems to correspond to an SVILC seen in Figs.~\ref{current}-\ref{currentSP}, and  
 to the cyclotron orbit explained in Sections~\ref{section8} and \ref{section9} for the conventional superconductor where electron-pairing occurs between band electrons.  It may be better to associate the ``idle wheel'' to the Berry connection whose non-trivial center locates at the center of the idle wheel.

A marked difference exists between `` idle wheels'' and `` topologically-protected loop currents''. While the idle wheels need to move to produce electric current, the topologically-protected loop currents do not have to. When the latter generate a current without translational motion, it is  a dissipationless current. The translational motion of the idle wheels and that of the topologically-protected loop currents both generate dissipative currents. They correspond to the current produced by the flow of the vortices in superconductors. The topologically-protected loop currents can appear and disappear, abruptly, but the idle wheels can not.

Lastly, we would like to emphasize that the relevant canonical transformation for the superfluid system is not the original Bogoliubov transformation that violates the particle number conservation, but the one that conserves the particle number \cite{koizumi2019}.
Using the particle number conserved version of the Bogoliubov transformation, we can explain the superfluid phenomena without the breakdown of the global $U(1)$ gauge invariance.
We expect that the new formalism will help to elucidate the cuprate superconductivity, where the local particle number fluctuation is highly suppressed by the strong correlation effect.

\bibliographystyle{ws-mplb}
\bibliography{SPIN-BCS}

\end{document}